\documentclass[notitlepage,showkeys,onecolumn,11pt,times,superscriptaddress, a4paper, pre, nofootinbib]{revtex4-2}
\usepackage[english]{babel}
\usepackage{stix}
\usepackage[T1]{fontenc}
\usepackage{graphicx}
\usepackage{bm} 
\usepackage[colorlinks=true,
            linkcolor=blue,
            urlcolor=blue,
            citecolor=blue]{hyperref}
\usepackage[mathlines]{lineno} 
\usepackage{amsmath}
\usepackage{physics}
\usepackage{booktabs}
\usepackage{geometry}
\usepackage{xcolor}
\usepackage{listings}

\definecolor{keywordcolor}{rgb}{0,0,0.6}
\definecolor{commentcolor}{rgb}{0.3,0.5,0.3}
\definecolor{stringcolor}{rgb}{0.6,0.1,0.1}
\definecolor{backgroundcolor}{rgb}{0.95,0.95,0.95}
\definecolor{framecolor}{rgb}{0.8,0.8,0.8}
\definecolor{promptcolor}{rgb}{0.5,0,0}  

\lstdefinestyle{python}{
  language=python,
  basicstyle=\ttfamily\small,
  numbers=left,
  numberstyle=\tiny\color{gray},
  stepnumber=1,
  numbersep=6pt,
  tabsize=4,
  breaklines=true,
  frame=single,
  rulecolor=\color{framecolor},
  keywordstyle=\color{keywordcolor}\bfseries,
  commentstyle=\color{commentcolor}\itshape,
  stringstyle=\color{stringcolor},
  showstringspaces=false,
  escapeinside={\%*}{*)},
  texcl=true,
  breaklines=true, breakatwhitespace=false, breakautoindent=true, postbreak=\mbox{\textcolor{gray}{$\hookrightarrow$}}
}

\lstdefinestyle{pycon}{
  language=python,
  basicstyle=\ttfamily\small,
  numbers=left,
  numberstyle=\tiny\color{gray},
  stepnumber=1,
  numbersep=6pt,
  tabsize=4,
  breaklines=true,
  frame=lines,
  rulecolor=\color{framecolor},
  keywordstyle=\color{keywordcolor}\bfseries,
  commentstyle=\color{commentcolor}\itshape,
  stringstyle=\color{stringcolor},
  showstringspaces=false,
  escapeinside={<@}{@>},
  texcl=true,
  alsoletter={>>>, >},
  morekeywords = {>>>}
}

\newcommand{\pyinline}[1]{\lstinline[language=python, basicstyle=\ttfamily\small]{#1}}

\lstdefinestyle{shell}{
  language=bash,
  basicstyle=\ttfamily\small,
  numbers=left,
  numberstyle=\tiny\color{gray},
  stepnumber=1,
  numbersep=6pt,
  breaklines=true,
  frame=single,
  rulecolor=\color{framecolor},
  keywordstyle=\color{keywordcolor}\bfseries,
  commentstyle=\color{commentcolor}\itshape,
  stringstyle=\color{stringcolor},
  texcl=true
}

\begin{document}
    \title{pynamicalsys: A Python toolkit for the analysis of dynamical systems}

    \author{Matheus Rolim Sales}
    \email{rolim.sales.m@gmail.com}
    \affiliation{University of Essex, School of Mathematics, Statistics and Actuarial Science, Wivenhoe Park, Colchester, CO4 3SQ, United Kingdom}
    \affiliation{São Paulo State University (UNESP), Institute of Geosciences and Exact Sciences, 13506-900, Rio Claro, SP, Brazil}
    \author{Leonardo Costa de Souza}
    \affiliation{Department of Physics, Institute for Complex Systems and Mathematical Biology, SUPA, University of Aberdeen, AB24 3UX, Aberdeen, United Kingdom}
    \affiliation{Institute of Physics, University of São Paulo, 05315-970, São Paulo, SP, Brazil}
    \author{Daniel Borin}
    \affiliation{São Paulo State University (UNESP), Institute of Geosciences and Exact Sciences, 13506-900, Rio Claro, SP, Brazil}
    \author{Michele Mugnaine}
    \affiliation{Institute of Physics, University of São Paulo, 05315-970, São Paulo, SP, Brazil}
    \author{José Danilo Szezech Jr.}
    \affiliation{Institute of Physics, University of São Paulo, 05315-970, São Paulo, SP, Brazil}
    \affiliation{Graduate Program in Science, State University of Ponta Grossa, 84030-900, Ponta Grossa, PR, Brazil}
    \affiliation{Department of Mathematics and Statistics, State University of Ponta Grossa, 84030-900, Ponta Grossa, PR, Brazil}
    \author{Ricardo Luiz Viana}
    \affiliation{Federal University of Paraná, Interdisciplinary Center for Science, Technology and Innovation, Center for Modeling and Scientific Computing, 81531-980, Curitiba, PR, Brazil}
    \author{Iberê Luiz Caldas}
    \affiliation{Institute of Physics, University of São Paulo, 05315-970, São Paulo, SP, Brazil}
    \author{Edson Denis Leonel}
    \affiliation{São Paulo State University (UNESP), Institute of Geosciences and Exact Sciences, 13506-900, Rio Claro, SP, Brazil}
    \author{Chris G. Antonopoulos}
    \affiliation{University of Essex, School of Mathematics, Statistics and Actuarial Science, Wivenhoe Park, Colchester, CO4 3SQ, United Kingdom}

    \begin{abstract}
        Since Lorenz's seminal work on a simplified weather model, the numerical analysis of nonlinear dynamical systems has become one of the main subjects of research in physics. Despite of that, there remains a need for accessible, efficient, and easy-to-use computational tools to study such systems. In this paper, we introduce \pyinline{pynamicalsys}, a simple yet powerful open-source Python module for the analysis of nonlinear dynamical systems. In particular, \pyinline{pynamicalsys} implements tools for trajectory simulation, bifurcation diagrams, Lyapunov exponents and several others chaotic indicators, period orbit detection and their manifolds, as well as escape and basins analysis. It also includes many built-in models and the use of custom models is straighforward. We demonstrate the capabilities of \pyinline{pynamicalsys} through a series of examples that reproduces well-known results in the literature while developing the mathematical analysis at the same time. We also provide the Jupyter notebook containing all the code used in this paper, including performance benchmarks. \pyinline{pynamicalsys} is freely available via the Python Package Index (PyPI) and is indented to support both research and teaching in nonlinear dynamics.
    \end{abstract}
    
    \date{\today}
    \maketitle

\section{Introduction}
    \label{sec:intro}

    The success of Newton's theory on mechanics led to the idea of a deterministic, and fully predictable Universe. Laplace once famously stated that if an intellect at a certain moment in time would know all the forces that set nature in motion, and all the positions of all objects, then this intellect would be able to predict the past and the future of the entire Universe~\cite{Laplace1826}. This idea was later challenged by Poincaré in his seminal work on the stability of the solar system~\cite{poincaré1893méthodes}. Poincaré, to simplify the problem, considered the gravitation interaction of only three objects, and demonstrated that the system is generally non-integrable. In other words, for an arbitrary initial condition, its motion cannot be described by a finite set of integrals of motion. Only a particular set of initial conditions results in exact solutions. He also discovered homoclinic points and homoclinic tangles, which make the motion sensitive to initial conditions and exhibit unpredictable behavior.

    This was the first evidence of the deterministic chaos, or simply chaos, that we know today. Poincaré, however, could not visualize the behavior he was describing. There were no computers at the time and his work remained underappreciated for around 70 years. In 1963, Lorenz when working on a simplified weather model, accidentally discovered that small changes in the initial conditions can lead to considerably different future states~\cite{Lorenz1963}. He considered a system of three differential equations and showed for the very first time a strange attractor: a geometric structure in phase space that is deterministic, i.e., follows a set of rules (the differential equations) yet is aperiodic and highly sensitive to small changes in the initial conditions.

    After Lorenz's discovery, the mathematical foundations of dynamical systems theory were revitalized. Smale \cite{Smale1963, Smale1967} introduced the concept of the horseshoe map, illustrating how repeated stretching and folding of phase space can lead to unpredictable behavior. Around the same time, the foundation of ergodic theory was also developed \cite{Sinai1963, Sinai1963b} and the term ``chaos'' started to become popular in the scientific community. The rapid development of computers and different programming languages, such as Pascal, Assembly, C, and Fortran, popular among the scientific community at the time, made large-scale numerical simulations feasible. In this context, the seminal work of Li and Yorke~\cite{Li1975} formalized the modern notion of chaos by showing that a system with a period-3 orbit must exhibit chaotic dynamics. Feigenbaum also contributed to this when he discovered a universal law in period-doubling bifurcation, today known as Feigenbaum constant, that shows that different systems can exhibit the same route to chaotic dynamics \cite{Feigenbaum1978, Feigenbaum1979}.

    Since then, and continuing to the present day, chaos theory has become a cornerstone of nonlinear science, with profound interdisciplinary influence. In physics, areas such as plasma physics~\cite{Chirikov1979, Morrison1980, Cary1983, Escande2016}, fluid turbulence~\cite{Chian2013, Ecke2015}, and astrophysical systems \cite{Contopoulos2002} rely heavily on chaotic models to describe their complex behavior. However, the reach of chaos theory extends far beyond physics. In biology, it plays a critical role in understanding heart rhythms~\cite{Lefebvre1993, Lombardi2000, Ferreira2011}, ecological models~\cite{Blasius2000, Dirk2009}, and neuronal activity~\cite{Labos1987, Rabinovich1997, Potapov2000, Guckenheimer2002}. The latter has become a major focus of current research. Chaotic models have also influenced the economics field by modeling complex market behavior and financial instabilities~\cite{peters1994fractal, Hibbert1994, Klioutchnikov2017}, while in computer science, it has influenced fields like cryptography~\cite{Baptista1998, Kocarev2001, kocarev2011chaos} and random number generation~\cite{Pareschi2006, Yu2019}.

    Therefore, in this paper, we present the \pyinline{pynamicalsys} module, a simple yet powerful, open-source Python package implementing several tools for the analysis of nonlinear dynamical systems. Despite being written purely in Python, \pyinline{pynamicalsys} offers high-performance thanks to Numba\footnote{\url{https://numba.pydata.org}}~\cite{numba} accelerated computation, offering speedups up to 130x compared to the pure Python version of the corresponding functions. We choose Python for its simplicity and extensive use within the scientific and programming communities, as it is currently one of the most, if not the most, widely used programming languages. You can install \pyinline{pynamicalsys} using the Python Package Index (PyPI) via
    \begin{lstlisting}[style=shell]
$ pip install pynamicalsys
    \end{lstlisting}

    We present the \pyinline{pynamicalsys} classes and methods and illustrate their usage together with the theoretical discussion of the methods. We use \pyinline{pynamicalsys} to reproduce several known results in the literature. This paper is accompanied by a Jupyter notebook (see the Supplementary Material), which contains all the code needed to reproduce the results presented in this paper. The notebook also shows the CPU time for each calculation, confirming the high efficiency of \pyinline{pynamicalsys}. These benchmarks were obtained on a MacBook Air equipped with an Apple M4 chip, featuring a 10-core CPU. In this paper, we only provide examples on how to obtain the data. For the plotting settings, we refer the reader to the Supplementary Material and the documentation page (\href{https://pynamicalsys.readthedocs.io/en/latest/}{https://pynamicalsys.readthedocs.io/en/latest/}).

    This paper is organized as follows. In Sec.~\ref{sec:basic}, we demonstrate the basic use of the \pyinline{DiscreteDynamicalSystem} class and perform some basic simulations such as trajectory and bifurcation diagram computation. In Sec.~\ref{sec:chaoticindicators}, we review some of the most used and efficient chaotic indicators for discrete dynamical systems and demonstrate their use by reproducing some known results in the literature. In Sec.~\ref{sec:manifolds}, we focus on finding and classifying periodic orbits of two-dimensional maps and determining the stable and unstable manifolds of the saddles. Section~\ref{sec:escape} is devoted to the escape analysis, such as the computation and quantification of escape basins (or attraction basins) and, lastly, Sec.~\ref{sec:concl} contains our final remarks.

    \section{Basic system definition and simulation}
    \label{sec:basic}

    We begin by presenting a few basic simulations on how to generate trajectories and phase space. We use the Chirikov-Taylor standard map~\cite{Chirikov1979}, defined as
    \begin{equation}
        \label{eq:stdmap}
        \begin{aligned}
            y_{n + 1} &= y_n + \frac{k}{2\pi}\sin\qty(2\pi x_n)\bmod1,\\
            x_{n + 1} &= x_n + y_{n + 1}\bmod1.
        \end{aligned}
    \end{equation}
    The Chirikov-Taylor standard map is a two-dimensional, area-preserving map where $x_n$ and $y_n$ are the conjugated canonical variables, $n = 0, 1, 2\ldots$, is the discrete time, and $k \geq 0$ is the nonlinearity parameter. For $k = 0$, the system is integrable and all orbits lie on period and quasiperiodic invariant tori. For $k > 0$, the sufficient irrational tori survive the perturbation, as predicted by the Kolmogorov-Arnold-Moser (KAM) theorem~\cite{lichtenberg2013regular} and the rational ones are destroyed, leaving behind a set of elliptic (center) and hyperbolic (saddle) periodic orbits (Poincaré-Birkhoff theorem~\cite{lichtenberg2013regular}). The stability islands are formed around the elliptic orbits and the stable and unstable manifolds of the hyperbolic orbits intersect each other in infinitely many points, generating chaotic dynamics. We refer the reader to Refs.~\cite{Chirikov1979,lichtenberg2013regular, Zaslavsky2000, Zaslavsky2002, Manos2014SurveyMap, Harsoula2019CharacteristicMap} for further details on the dynamics of the standard map.

\begin{figure}[t]
        \centering
        \includegraphics[width=\linewidth]{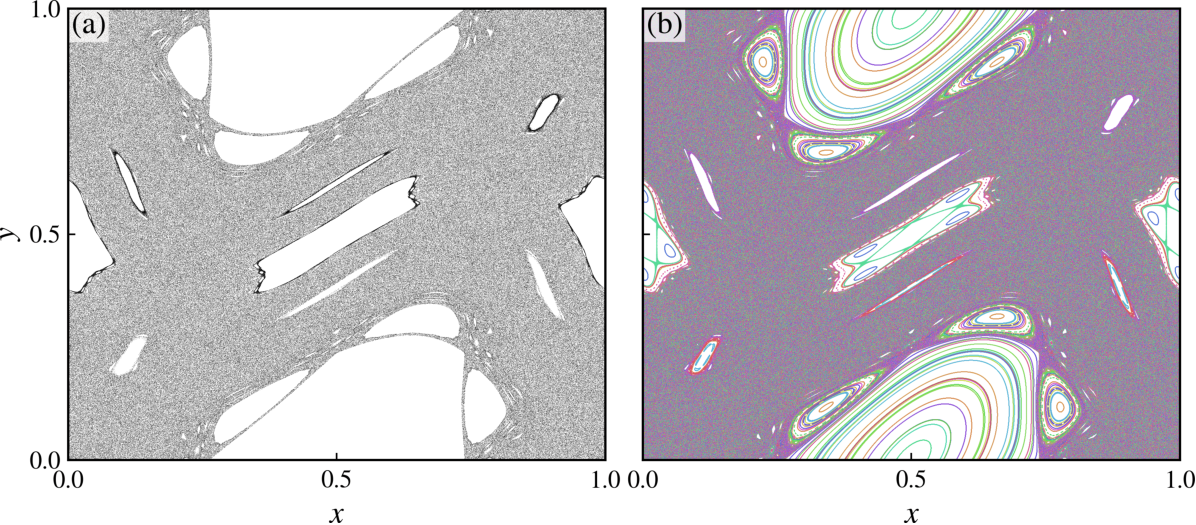}
        \caption{Demonstration of the use of the \pyinline{trajectory} method from the \pyinline{DiscreteDynamicalSystem} class of \pyinline{pynamicalsys} for the \pyinline{model="standard map"}, with parameters $k = 1.5$ and total\textunderscore time = $100000$, for (a) a single initial condition and (b) $200$ initial conditions.}
        \label{fig:fig1}
    \end{figure}

    Let us then create our dynamical system object using the \pyinline{DiscreteDynamicalSystem} class. To import the class, we proceed as follows:
    \begin{lstlisting}[style=pycon]
>>> from pynamicalsys import DiscreteDynamicalSystem as dds
    \end{lstlisting}
    The class takes on six arguments: \pyinline{model}, \pyinline{mapping}, \pyinline{jacobian}, \pyinline{backwards_mapping}, \pyinline{system_dimension}, and \pyinline{number_of_parameters}. You should either inform \pyinline{model} \textit{or} the remaining five arguments. The \pyinline{DiscreteDynamicalSystem} class comes with a few built-in systems and the standard map is one of them. To check all the built-in systems, run
    \begin{lstlisting}[style=pycon]
>>> dds.available_models()
['standard map', 'unbounded standard map', 'henon map', 'lozi map', 'rulkov map', 'logistic map', 'standard nontwist map', 'extended standard nontwist map', 'leonel map', '4d symplectic map']
    \end{lstlisting}
    Thus, to create an object of the standard map system, you proceed as
    \begin{lstlisting}[style=pycon]
>>> ds = dds(model="standard map")
    \end{lstlisting}
    Now all methods of the \pyinline{DiscreteDynamicalSystem} class are accessible via the \pyinline{ds} object. To generate the trajectories, we use the \pyinline{trajectory} method:
    \begin{lstlisting}[style=python]
obj.trajectory(u, total_time, parameters=None, transient_time=None)
    \end{lstlisting}
    It takes on four arguments: \pyinline{u}, the initial condition, \pyinline{total_time} defines the total iteration time, \pyinline{parameters} is a list of the system parameters, which can be left empty if the system has no parameters, and \pyinline{transient_time} corresponds to the discarded initial iterations (default is \pyinline{None}). The initial condition can be an one-dimensional array, corresponding to a single initial condition [Fig.~\ref{fig:fig1}(a)]:
    \begin{lstlisting}[style=pycon]
>>> u = np.array([0.05, 0.05])  # Initial condition
>>> k = 1.5  # Parameter of the map
>>> total_time = 1000000  # Total iteration time
>>> trajectory = ds.trajectory(u, total_time, parameters=k)
>>> trajectory.shape
(1000000, 2)
    \end{lstlisting}
    In this case, the \pyinline{trajectory()} method returns a two-dimensional array with \pyinline{total_time} rows (the state at each time step) and $2$ columns (the $x$ and $y$ coordinates). The initial condition \pyinline{u} can also be a two-dimensional array of shape $(M, 2)$, where $M$ is the number of initial conditions and $2$ corresponds to the dimension of the system [Fig.~\ref{fig:fig1}(b)]:
    \begin{lstlisting}[style=pycon]
>>> num_ic = 200  # Number of initial conditions
>>> np.random.seed(13)  # For reproducibility
>>> u = np.random.rand(num_ic, 2)  # Random initial conditions
>>> k = 1.5  # Parameter of the map
>>> total_time = 100000  # Total iteration time for each initial condition
>>> trajectories = ds.trajectory(u, total_time, parameters=k)
>>> trajectories.shape
(20000000, 2)
    \end{lstlisting}
    Now, the \pyinline{trajectory} method returns a two-dimensional array with \pyinline{total_time}$\,\times\,$\pyinline{num_ic} rows, i.e., it concatenates each trajectory and returns the trajectory of all initial conditions. Figures~\ref{fig:fig1}(a) and (b) present complementary and key features of the standard map. In case (a), a single chaotic initial condition was used. This illustrates the chaotic sea, the boundaries of multiple islands that act as barriers in phase space (represented by the white areas), and, finally, the stickiness around the central islands, evidenced by the concentration of points indicating that the trajectory becomes temporarily trapped for long but finite times~\cite{Zaslavsky2000,Sales2023,Souza2024}. On the other hand, case (b) demonstrates how easily \pyinline{pynamicalsys} can be adjusted to run multiple initial conditions. Each initial condition is marked with a distinct color, clearly distinguishing the regular regions, represented by islands, from the chaotic sea that fills the phase space.

\begin{figure}[t]
        \centering
        \includegraphics[width=\linewidth]{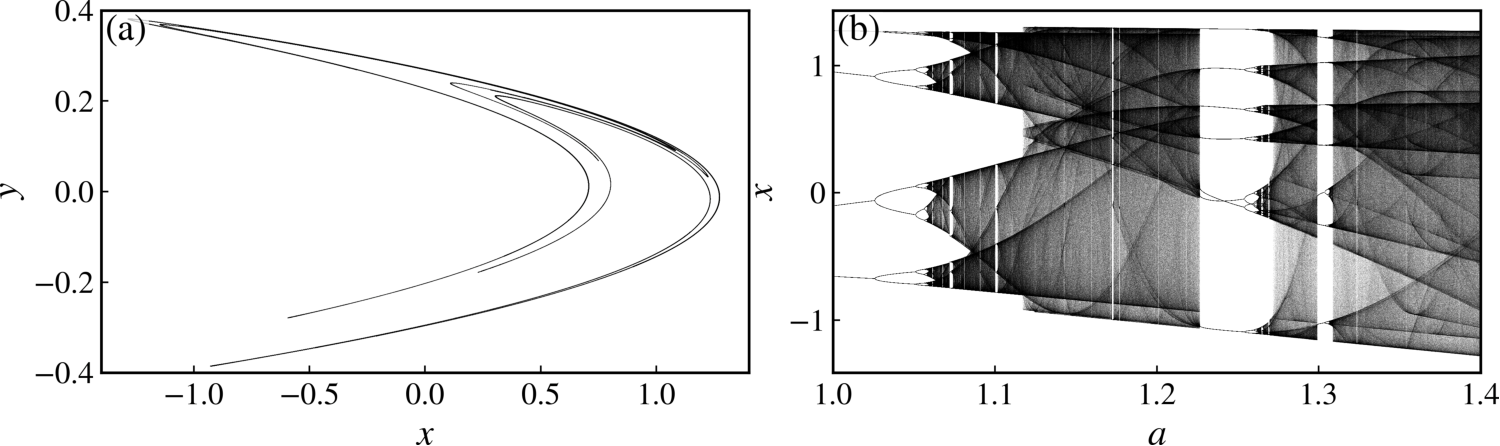}
        \caption{Demonstration of the use of the \pyinline{trajectory} method from the \pyinline{DiscreteDynamicalSystem} class in \pyinline{pynamicalsys} for the \pyinline{model="henon map"} with parameters $a = 1.4$ and $b = 0.3$ in (a), and of the \pyinline{bifurcation_diagram} method over the interval $a \in [1.0, 1.4]$ with $b = 0.3$ in (b).}
        \label{fig:fig2}
    \end{figure}

    Let us now consider a two-dimensional, dissipative map: the Hénon map~\cite{henon1976}. The map is defined as
    \begin{equation}
        \label{eq:henonmap}
        \begin{aligned}
            x_{n + 1} &= 1 - ax_n^2 + y_n,\\
            y_{n + 1} &= b x_n,
        \end{aligned}
    \end{equation}
    where $a$ and $b$ are the parameters of the system. This system is built-in within the \pyinline{DiscreteDynamicalSystem} class as well. To initialize it we simply define the dynamical system object with \pyinline{model="henon map"}. In this case, we are dealing with two parameters instead of one, and the order in which they are passed to the methods matters. To obtain this information from the built-in systems, you proceed as follows:
    \begin{lstlisting}[style=pycon]
>>> from pynamicalsys import DiscreteDynamicalSystem as dds
>>> ds = dds(model="henon map")
>>> info = ds.info
>>> info["parameters"]
['a', 'b']
    \end{lstlisting}
    The \pyinline{info} property returns a dictionary with several information regarding the built-in system. For all the available information, check the documentation. Thus, from the previous example, we see that for the built-in Hénon map, the first parameter is $a$ and the second is $b$. Now, let us generate the chaotic Hénon attractor for the parameters $a = 1.4$ and $b = 0.3$ [Fig.~\ref{fig:fig2}(a)]:
    \begin{lstlisting}[style=pycon]
>>> from pynamicalsys import DiscreteDynamicalSystem as dds
>>> ds = dds(model="henon map")
>>> u = [0.1, 0.1]  # Initial condition
>>> a, b = 1.4, 0.3  # Parameters for the Henon map
>>> parameters = [a, b]
>>> total_time = 1000000  # total iteration time
>>> transient_time = 500000  # Transient time for the Henon map
>>> trajectory = ds.trajectory(u, total_time, parameters=parameters, transient_time=transient_time)
>>> trajectory.shape
(500000, 2)
    \end{lstlisting}
    The \pyinline{transient_time} argument tells the \pyinline{trajectory()} method the number of iterations to discard, thus it returns a two-dimensional array of \pyinline{total_time}$\,-\,$\pyinline{transient_time} rows.
    
    When studying dissipative systems, one useful tool is the bifurcation diagram. The dynamical systems, in general, depend on different parameters and as these parameters change, the system can undergo transitions known as bifurcations. We use the \pyinline{bifurcation_diagram} method of the \pyinline{DiscreteDynamicalSystem} class of \pyinline{pynamicalsys}:
    \begin{lstlisting}[style=python]
obj.bifurcation_diagram(u, param_index, param_range, total_time, parameters=None, transient_time=None, continuation=False, return_last_state=False, observable_index=0)
    \end{lstlisting}
    Here, \pyinline{u} is the initial condition, \pyinline{param_index} is an integer that corresponds to the position of the bifurcation parameter in the parameter list, i.e., \pyinline{0} is the first parameter, \pyinline{1} is the second, and so on. \pyinline{param_range} determines the bifurcation parameter values. It can either be a predefined sample of parameters, i.e., \pyinline{param_range = [0.1, 0.2, 0.3]} or it can be a tuple indicating the starting and ending values and the number of values for a linear spacing: \pyinline{param_range = (start, end, num_params)}. \pyinline{total_time} is the total iteration time, including the transient time, and \pyinline{parameters} is a list with the remaining parameter values, i.e., those that remain fixed. If the system only has one parameter, when computing the bifurcation diagram this should be set to \pyinline{None}. \pyinline{transient_time} is the initial iteration time to discard (default is \pyinline{None}) and \pyinline{continuation} determines whether to reset or not the initial condition for every new parameter value. If \pyinline{continuation=False}, for every new parameter value, the bifurcation diagram is computed using the provided initial condition \pyinline{u}. However, if \pyinline{continuation=True}, then it is performed a numerical continuation sweep, i.e., the initial condition for the next parameter value is the last state of the previous parameter (default is \pyinline{False}). The argument \pyinline{return_last_state} determines whether to return the last state as well (default is \pyinline{False}) and \pyinline{observable_index} corresponds to the coordinate used in the bifurcation diagram. By default, it uses the first coordinate. To use a different one, set this argument to \pyinline{observable_index=1}, to use the second coordinate, for example.

    For the Hénon map, we choose to fix $b = 0.3$ and we want to change $a$ in the interval $a \in[1.0, 1.4]$. Thus, we proceed as follows [Fig.~\ref{fig:fig2}(b)]:
    \begin{lstlisting}[style=pycon]
>>> from pynamicalsys import DiscreteDynamicalSystem as dds
>>> ds = dds(model="henon map")
>>> u = [0.1, 0.1]  # Initial condition
>>> b = 0.3  # Keep b fixed
>>> parameters = b
>>> param_range = (1, 1.4, 2500)  # Parameter range (a in this case)
>>> param_index = 0  # a is going to be changed (parameters = [a, b])
>>> total_time = 8000  # Total iteration time
>>> transient_time = 2000  # Transient time
>>> param_values, bifurcation_diagram = ds.bifurcation_diagram(u, param_index, param_range, total_time, parameters=parameters, transient_time=transient_time)
>>> bifurcation_diagram.shape
(2500, 6000)
    \end{lstlisting}
    The \pyinline{bifurcation_diagram()} method returns two arrays: the first contains all parameter values, and the second is a 2D array representing the corresponding coordinates for each parameter value. If \pyinline{return_last_state} is set to \pyinline{True}, then it also returns the last state of the system at the final parameter value. This can be useful when studying multistability and hysteresis, intriguing phenomena that can be present in diverse dynamical systems.~\cite{FEUDEL2008,PISARCHIK2014,Mugnaine2022,Bi2024}.

    \section{Chaotic indicators}
    \label{sec:chaoticindicators}
    
    Chaos is notably present in both natural phenomena and mathematical models. Developing reliable methods and tools to identify and distinguish between chaotic and periodic behaviors is often a crucial objective. Numerous numerical techniques, each with distinct approaches, have been proposed to detect the presence of chaotic dynamics. To facilitate this, the \pyinline{DiscreteDynamicalSystem} class of \pyinline{pynamicalsys} offers a variety of easy-to-use methods.
    
    This section is organized as follows: (i) Lyapunov exponents, (ii) Linear dependence indexes, (iii) Weighted Birkhoff averages, (iv) Recurrence time entropy, and (v) Hurst exponent.

    \subsection{Lyapunov exponents}
    \label{subsec:LE}

    The Lyapunov exponents are a measure of the average exponential rates at which infinitesimal perturbations to a trajectory in a dynamical system grow or decay over time. Given an one-dimensional discrete dynamical system $x_{n + 1} = f(x_n)$, where $f:\mathbb{R} \rightarrow \mathbb{R}$ is a smooth map. Let $x_0$ and $y_0 = x_0 + \delta_0$ be two initial conditions infinitesimally close to each other ($\abs{\delta_0} \ll 1$). After one iteration, the two initial conditions become
    \begin{equation}
        \begin{aligned}
            x_1 &= f(x_0),\\
            y_1 &= f(x_0 + \delta_0).
        \end{aligned}
    \end{equation}
    The difference between them is
    \begin{equation}
        \delta_1 = y_1 - x_1 = f(x_0 + \delta_0) - f(x_0).
    \end{equation}
    For $\delta_0$ small enough, we can linearize $f(x_0 + \delta_0)$ using a first-order Taylor expansion:
    \begin{equation}
        f(x_0 + \delta_0) = f(x_0) + f'(x_0)\delta_0 + \mathcal{O}(\delta_0^2),
    \end{equation}
    leading us to $\delta_1 = f'(x_0)\delta_0$. For the next iteration, we obtain $\delta_2 = f'(x_1)f'(x_0)\delta_0$. Thus, repeating this for $n$ iterations, we obtain
    \begin{equation}
        \delta_n = \qty(\prod_{i = 0}^{n - 1}f'(x_i))\delta_0.
    \end{equation}
    We want to quantify how this perturbation grows exponentially. Thus, we derive the exponential rate of divergence:
    \begin{equation}
        \frac{1}{n}\log\abs{\frac{\delta_n}{\delta_0}} = \frac{1}{n}\log\qty(\prod_{i = 0}^{n - 1}\abs{f'(x_i)}) = \frac{1}{n}\sum_{i = 0}^{n - 1}\log\abs{f'(x_i)}.
    \end{equation}
    The Lyapunov exponent $\lambda$ is defined as the long-term average exponential rate of separation, i.e.,
    \begin{equation}
        \label{eq:lypnv1d}
        \lambda = \lim_{n\rightarrow\infty}\frac{1}{n}\sum_{i = 0}^{n - 1}\log\abs{f'(x_i)}.
    \end{equation}

    For higher-dimensional dynamical systems, the derivation is somewhat different. Given a $d$-dimensional discrete-time dynamical system $\vb{x}_{n + 1} = \vb{f}(\vb{x}_n)$, where $\vb{f}:\mathbb{R}^d\rightarrow\mathbb{R}^d$ is a smooth map, the Lyapunov spectrum $\lambda_1 \geq \lambda_2 \geq \ldots \geq \lambda_d$ is defined under Oseledec's multiplicative ergodic theorem~\cite{Ott2002}. Let $J(\vb{x}) = \vb{Df}(\vb{x})$ be the Jacobian matrix of the map $\vb{f}$ at point $\vb{x}$. The matrix
    \begin{equation}
        \label{eq:prodjacs}
        J_n(\vb{x}_0) = J(\vb{x}_{n - 1})J(\vb{x}_{n - 2})\ldots J(\vb{x}_{1})J(\vb{x}_{0})
    \end{equation}
    describes the evolution of the tangent vectors under the linearized dynamics. Oseledec's theorem states that for almost every initial condition $\vb{x}_0$, the following limit exists:
    \begin{equation}
        \label{eq:Lambda}
        \Lambda(\vb{x}_0) = \lim_{n\rightarrow\infty}\qty[J_n^T(\vb{x}_0)J_n(\vb{x}_0)]^{1/2n}.
    \end{equation}
    The Lyapunov exponents are related to the eigenvalues of the matrix $\Lambda$ and are given by
    \begin{equation}
        \label{eq:lyapunov}
        \lambda_i = \lim_{n\rightarrow\infty}\frac{1}{n}\log\norm{J_n(\vb{x}_0)\vb{v}_i},
    \end{equation}
    where $\vb{v}_i$ are the corresponding eigenvector of $\Lambda(\vb{x}_0)$.

    While Eqs.~\eqref{eq:Lambda} and~\eqref{eq:lyapunov} are extremely elegant in a theoretical sense, directly computing the matrix $J_n(\vb{x}_0)$ leads to numerical instability, and the product becomes dominated by the direction of maximal growth, making it impossible to calculate the smaller Lyapunov exponents. To overcome this issue, several methods have been proposed to numerically estimate the Lyapunov exponents~\cite{Shimada1979, Benettin1980, Wolf1985, Eckmann1985}. We describe in the following the QR-based approach that has become the standard procedure in numerical studies of the Lyapunov exponents. The idea of this method is to evolve an orthonormal basis of tangent vectors and reorthonormalize them at each step using a QR decomposition. Given an initial orthonormal matrix $Q_0 \in \mathbb{R}^{d\times d}$, where each column of $Q_0$ is a tangent vector, the evolution of $Q_0$ according to the linearized dynamics is
    \begin{equation}
        A_1 = J_1Q_0.
    \end{equation}
    We now perform a QR decomposition on $A_1$: $A_1 = Q_1R_1$, where $Q_1 \in \mathbb{R}^{d\times d}$ is an orthonormal matrix and $R_1 \in \mathbb{R}^{d\times d}$ is an upper triangular matrix. We can then write
    \begin{equation}
        J_1 = Q_1R_1Q_0^{-1}.
    \end{equation}
    For the second iteration, the procedure is analogous and we obtain $J_2 = Q_2R_2Q_1^{-1}$. By repeating this procedure recursively, we can express the Jacobian matrix at each time $n$ as
    \begin{equation}
        \label{eq:jn}
        J_n = Q_nR_nQ_{n - 1}^{-1}.
    \end{equation}
    
    The physical meaning behind this method is as follows: the Jacobian matrix $J_n$ evolves the orthonormal basis $Q_{n - 1}$ under the linearized dynamics, resulting in a new set of vectors $A_n = J_nQ_{n - 1}$, which are generally not orthonormal. If we continually evolve this set of vectors, eventually all of them align with the direction of maximal growth and we lose all information about the other Lyapunov exponents. When we perform a QR decomposition on $A_n$, yielding $A_n = Q_nR_n$, we are reorthonormalizing the tangent basis to prevent the collapse onto the direction of maximum growth, while simultaneously extracting the local stretching and shrinking information captured by the matrix $R_n$. The matrix $Q_n$ then becomes the updated orthonormal basis, aligned with the principal directions of local deformation. The absolute values of the diagonal elements of $R_n$, $\abs*{r_{ii}^{(n)}}$, represent the instantaneous rates of expansion or contraction along each orthonormal direction. Therefore, by computing the time averages of the logarithms of these values, we can estimate the Lyapunov exponents:
    \begin{equation}
        \label{eq:lyapunovR}
        \lambda_i = \lim_{n\rightarrow\infty}\frac{1}{n}\sum_{j = 1}^n\log{\abs{r_{ii}^{(j)}}}.
    \end{equation}

    This QR method is not just a computational technique. It is connected with the theoretical framework established by Oseledec's theorem [Eq.~\eqref{eq:Lambda}]. To see that, let us substitute each Jacobian matrix in terms of its QR decomposition in the matrix product in Eq.~\eqref{eq:prodjacs}. We get
    \begin{equation}
        \begin{aligned}
            J_n = \qty(Q_nR_nQ_{n - 1}^{-1})\qty(Q_{n - 1}R_{n - 1}Q_{n - 2}^{-1})\ldots\qty(Q_1R_1Q_{0}^{-1}).
        \end{aligned}
    \end{equation}
    All of the $Q_k^{-1}$ and $Q_k$ cancel due to the orthogonality of the matrices and the expression simplifies to
    \begin{equation}
        J_n = Q_n\qty(R_nR_{n - 1}\ldots R_1)Q_0^{-1} = Q_n\mathcal{R}_nQ_0^{-1},
    \end{equation}
    where $\mathcal{R}_n = R_nR_{n - 1}\ldots R_1$. Now, we compute the product $J_n^TJ_n$:
    \begin{equation}
        \begin{aligned}
            J_n^TJ_n &= (Q_0^{-1})^T\mathcal{R}_n^TQ_n^TQ_n\mathcal{R}_nQ_0^{-1},\\
            &= Q_0^{-T}\mathcal{R}_n^T\mathcal{R}_nQ_0^{-1}.
        \end{aligned}
    \end{equation}
    This construction leads to a similarity transformation: the matrix $\mathcal{R}_n^T\mathcal{R}_n$ is similar to $J_n^TJ_n$, and thus both share the same eigenvalues. As $n \rightarrow \infty$, the eigenvalues of the matrix $J_n^TJ_n$ converge to those of the limiting matrix $\Lambda(\vb{x}_0)$ [Eq.~\eqref{eq:Lambda}], whose eigenvalues define the Lyapunov exponents. Therefore, the eigenvalues of $\mathcal{R}_n^T\mathcal{R}_n$ are a numerical approximation to those of $J_n^TJ_n$ and consequently yield the same Lyapunov exponents in the long-time limit [Eq.~\eqref{eq:lyapunovR}].
    
    For two-dimensional systems, it is possible to derive an analytical recursive expression for the diagonal elements of $R_n$. We can rewrite Eq.~\eqref{eq:jn} as
    \begin{equation}
        \label{eq:rn}
        R_n = Q_n^{-1}J_nQ_{n - 1}.
    \end{equation}
    Let us choose as our orthogonal matrix the two-dimensional rotation matrix:
    \begin{equation}
        Q_{n} = \mqty(\cos\beta_n & -\sin\beta_n \\ \sin\beta_n & \cos\beta_n).
    \end{equation}
    This matrix rotates a two-dimensional vector counterclockwise by an angle $\beta$. Thus, Eq.~\eqref{eq:rn} becomes
    \begin{equation}
        R_n = \mqty(r_n^{(11)} & r_n^{(12)} \\ 0 & r_n^{(22)}) = \mqty(\cos\beta_n & \sin\beta_n \\ -\sin\beta_n & \cos\beta_n)\mqty(J_n^{(11)} & J_n^{(12)} \\ J_n^{(21)} & J_n^{(22)})\mqty(\cos\beta_{n - 1} & -\sin\beta_{n - 1} \\ \sin\beta_{n - 1} & \cos\beta_{n - 1})
    \end{equation}
    The diagonal elements are then given by
    \begin{equation}
        \label{eq:rndiag}
        \begin{aligned}
            r_n^{(11)} &= \cos\beta_n \qty(J_n^{(11)} \cos\beta_{n-1} + J_n^{(12)} \sin\beta_{n-1}) + \sin\beta_n\qty(J_n^{(21)} \cos\beta_{n-1} + J_n^{(22)} \sin\beta_{n-1}),\\
            r_n^{(22)} &= -\sin\beta_n \qty(-J_n^{(11)} \sin\beta_{n-1} + J_n^{(12)} \cos\beta_{n-1}) + \cos\beta_n \qty(-J_n^{(21)} \sin\beta_{n-1} + J_n^{(22)} \cos\beta_{n-1}),
        \end{aligned}
    \end{equation}
    and the relation between $\beta_n$ and $\beta_{n - 1}$ is
    \begin{equation}
        \label{eq:betan}
        \tan \beta_n = \frac{J_n^{(21)} \cos\beta_{n-1} + J_n^{(22)} \sin\beta_{n-1}}{J_n^{(11)} \cos\beta_{n-1} + J_n^{(12)} \sin\beta_{n-1}}.
    \end{equation}
    Therefore, given an initial angle $\beta_0$ (typically $\beta_0 = 0$), we first calculate $\beta_1$ using Eq.~\eqref{eq:betan}. Then, we calculate the diagonal elements of the matrix $R_n$ via Eq.~\eqref{eq:rndiag}. By repeating this process iteratively, we can calculate the Lyapunov exponents using Eq.~\eqref{eq:lyapunovR}.

    To calculate the Lyapunov exponents, we use the \pyinline{lyapunov} method of the \pyinline{DiscreteDynamicalSystem} class of \pyinline{pynamicalsys}:
    \begin{lstlisting}[style=python]
obj.lyapunov(u, total_time, parameters=None, method="QR", return_history=False, sample_times=None, transient_time=None, log_base=np.e) 
    \end{lstlisting}
    The arguments \pyinline{u}, \pyinline{total_time}, and \pyinline{parameters} are the initial condition, the list of parameters of the system, and the total iteration time, respectively. For an one-dimensional system, the \pyinline{lyapunov} method computes the Lyapunov exponent via Eq.~\eqref{eq:lypnv1d}. For more than one dimension, the optional argument \pyinline{method} determines which QR decomposition to use. For two-dimensional systems, if \pyinline{method="QR"}, it automatically uses the analytical recursive expression we have derived [Eqs.~\eqref{eq:rndiag} and~\eqref{eq:betan}]. For higher-dimensional systems, the default method employs the modified Gram-Schmidt algorithm to perform QR decomposition. If the problem requires improved numerical stability (e.g. very large-scale problem), it is possible to set \pyinline{method="QR_HH"} to use Householder reflections instead. The optional argument \pyinline{return_history}, when set to \pyinline{True} (default is \pyinline{False}) tells the `\pyinline{lyapunov} method to return the value of each Lyapunov exponent at every time step. However, for long simulations, this can result in extremely large arrays that may exceed the system's memory capacity. To avoid this, you can pass to the optional argument \pyinline{sample_times} a list of time points so that the \pyinline{lyapunov} method returns the Lyapunov exponents only at those specified times. The optional argument \pyinline{transient_time} determines the number of iterations to discard before starting the Lyapunov exponents calculation and \pyinline{log_base} is the base of the logarithm used to calculate the Lyapunov exponents (default is set to \pyinline{np.e}, i.e., the natural logarithm).

    Let us then consider a system that is not built-in. We consider the dissipative asymmetric kicked rotor map (DAKRM), defined as~\cite{Wang2007, Celestino2011, Lopes2012, RolimSales2024}
    \begin{equation}
        \label{eq:dakrm}
        \begin{aligned}
            y_{n + 1} &= (1 - \gamma)y_n + k\left[\sin(x_n) + a \sin\left(2x_n + \frac{\pi}{2}\right)\right],\\
            x_{n + 1} &= x_n + y_{n + 1}\bmod2\pi,
        \end{aligned}
    \end{equation}
    where $k \geq 0$ corresponds to the strength of the perturbation, similar to $k$ in the standard map [cf.~Eq.~\eqref{eq:stdmap}], $a$ is the asymmetry parameter which breaks the spatial symmetry for $a \neq 0$ and $\gamma \in[0, 1]$ is the dissipative parameter. For $a = 0$ and $\gamma = 0$, we recover the standard map. To calculate the Lyapunov exponents, we need the Jacobian matrix of the system, which is given by
    \begin{equation}
        J = \mqty(1 + \pdv{y_{n+1}}{x_n} & \pdv{y_{n+1}}{y_n} \\ \pdv{y_{n+1}}{x_n} & \pdv{y_{n+1}}{y_n}),
    \end{equation}
    where
    \begin{equation}
        \begin{aligned}
            \pdv{y_{n+1}}{x_n} &= k\qty[\cos(x_n) + 2 a\cos\qty(2x_n + \frac{\pi}{2})],\\
            \pdv{y_{n+1}}{y_n} &= 1 - \gamma.
        \end{aligned}
    \end{equation}
    Since the DAKRM is not built-in within the \pyinline{pynamicalsys} package, we need to define it ourselves. The mapping function signature should be \pyinline{u1 = f(u0, parameters)}, i.e., given the initial condition and the parameters, the function returns the next state:
    \begin{lstlisting}[style=pycon]
>>> from numba import njit
>>> @njit
>>> def dakrm(u, parameters):
...    k, a, gamma = parameters
...    x, y = u
...    y_new = (1 - gamma) * y + k * (np.sin(x) + a * np.sin(2 * x + np.pi / 2))
...    x_new = (x + y_new) % (2 * np.pi)
...    return np.array([x_new, y_new])
\end{lstlisting}

    Note the use of the \pyinline{@njit} decorator before the function definition. It is absolutely crucial that both the model and Jacobian functions are decorated with \pyinline{@njit}. This decorator enables Numba to compile the function to optimized machine code, resulting in a significant performance boost for numerical computations. Furthermore, since all methods within \pyinline{pynamicalsys} are also decorated with \pyinline{@njit}, the model functions must be decorated similarly to ensure compatibility. Without this, Numba would raise errors due to mixing compiled and uncompiled functions. Having defined the model function, the next step is to define the Jacobian matrix function. The Jacobian function signature is \pyinline{J = jac(u, parameters, *args)}, i.e., given the current state of the system and its parameters, the function returns a two-dimensional array of shape \pyinline{(d, d)}, where \pyinline{d} is the dimension of the system:
    \begin{lstlisting}[style=pycon]
>>> @njit
>>> def dakrm_jacobian(u, parameters, *args):
...    k, a, gamma = parameters
...    x, y = u
...    dFdx = k * (np.cos(x) + 2 * a * np.cos(2 * x + np.pi / 2))
...    dFdy = 1 - gamma
...    return np.array([
...        [1 + dFdx, dFdy],
...        [dFdx,     dFdy]])
    \end{lstlisting}

    Finally, with both the map equation and Jacobian matrix defined, we can create the dynamical system object as
    \begin{lstlisting}[style=pycon]
>>> ds = dds(mapping=dakrm, jacobian=dakrm_jacobian, system_dimension=2, number_of_parameters=3)
    \end{lstlisting}
    Even though \pyinline{dds} also takes on the \pyinline{backwards_mapping} argument, since we are not using it to calculate the Lyapunov exponents, it is not necessary to provide it. If, however, one wishes to calculate the manifolds of a dynamical system, for instance, then it becomes necessary. The point is, when defining a dynamical system that is not built-in, you only need to provide the functions that are going to be used. If you are only interested in drawing the trajectories, then you do not have to provide the Jacobian matrix, for instance.

    To illustrate the calculation of the Lyapunov exponents, let us fix $k = 8.0$ and $\gamma = 0.8$ and calculate them for $a = 0.47$:
    \begin{lstlisting}[style=pycon]
>>> u = [1.78, 0.0]  # Initial condition
>>> total_time = 10000 # Total iteration time
>>> transient_time = 5000  # Transient time
>>> k, a, gamma = 8, 0.47, 0.8  # Parameters of the system
>>> parameters = [k, a, gamma]
>>> ds.lyapunov(u, total_time, parameters=parameters, transient_time=transient_time)
array([-0.35202562, -1.25741229])
    \end{lstlisting}
    and for $a = 0.6$:
 \begin{lstlisting}[style=pycon]
>>> u = [1.78, 0.0]  # Initial condition
>>> total_time = 10000  # Total iteration time
>>> transient_time = 5000  # Transient time
>>> k, a, gamma = 8, 0.6, 0.8  # Parameters of the system
>>> parameters = [k, a, gamma]
>>> ds.lyapunov(u, total_time, parameters=parameters, transient_time=transient_time)
array([ 1.57224186, -3.18167977])
 \end{lstlisting}
    Thus, the \pyinline{lyapunov()} method returns an one-dimensional array with all the Lyapunov exponents in decreasing order. For $a = 0.47$ we obtain two negative Lyapunov exponents, indicating periodic dynamics, and for $a = 0.6$ the largest Lyapunov exponent is positive while the second one is negative, i.e., the dynamics is chaotic.

    \begin{figure}[t]
        \centering
        \includegraphics[width=\linewidth]{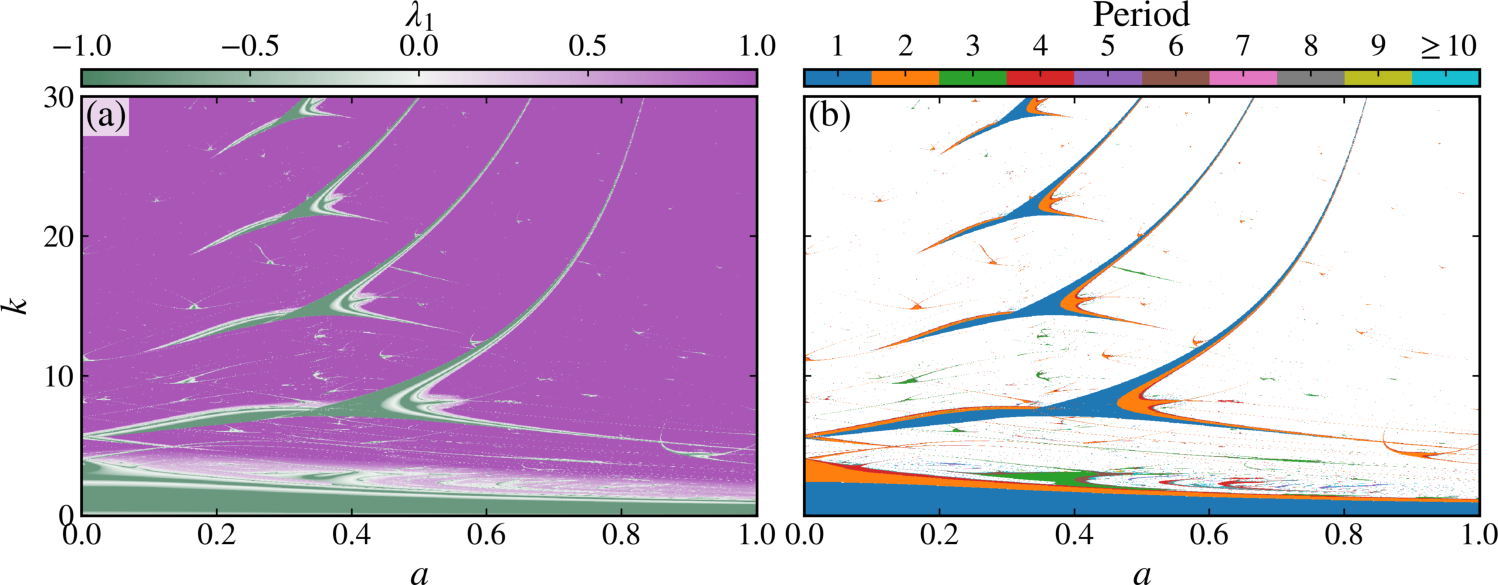}
        \caption{Demonstration of the use of the (a) \pyinline{lyapunov} method and (b) \pyinline{period} method from the \pyinline{DiscreteDynamicalSystem} class of \pyinline{pynamicalsys} for the dissipative kicked rotor map [Eq.~\eqref{eq:dakrm}] with parameter $\gamma = 0.8$.}
        \label{fig:fig3}
    \end{figure}
    
    Let us suppose now that you do not know the Jacobian matrix of your system. You can instantiate the \pyinline{dds} class without providing a Jacobian matrix function:
    \begin{lstlisting}[style=pycon]
>>> ds = dds(mapping=dakrm, system_dimension=2, number_of_parameters=3)
    \end{lstlisting}
    In this case, the Jacobian matrix is numerically determined as
    \begin{equation}
        J_{ij} = \frac{f_i(x_1, x_2, \ldots, x_j + \epsilon, \ldots, x_n) - f_i(x_1, x_2, \ldots, x_j - \epsilon, \ldots, x_n)}{2\epsilon},
    \end{equation}
    where $f_i$ is the $i$th component of the mapping and $\epsilon$ is chosen as
    \begin{equation}
        \epsilon = \left( \varepsilon_{\text{mach}} \right)^{1/3} \cdot \max\left(1, \|\mathbf{u}\|_2\right),
    \end{equation}
    where $\varepsilon_{\text{mach}}$ is the machine epsilon (the smallest representable difference between floating point numbers), $\vb{u} = (x_1, x_2, \ldots, x_d)^T$ is the state vector, and $\|\vb{u}\|_2$ is the Euclidean norm of the state vector. This choice is a compromise between truncation and round-off errors~\cite{press2007numerical}. The Lyapunov exponents is computed in the same way as before:
     \begin{lstlisting}[style=pycon]
>>> u = [1.78, 0.0]  # Initial condition
>>> total_time = 10000  # Total iteration time
>>> transient_time = 5000  # Transient time
>>> k, a, gamma = 8, 0.6, 0.8  # Parameters of the system
>>> parameters = [k, a, gamma]
>>> ds.lyapunov(u, total_time, parameters=parameters, transient_time=transient_time)
array([ 1.5740678 , -3.18114158])
    \end{lstlisting}
    The final value, of course, differs from the one obtained when we instantiated the \pyinline{dds} class with the Jacobian matrix.
    
    While the Lyapunov exponents tell us that for $a = 0.4$ the dynamics is periodic, they tell us nothing about the period itself. This information can be obtained using the \pyinline{period} of the \pyinline{DiscreteDynamicalSystem} class of \pyinline{pynamicalsys}:
    \begin{lstlisting}[style=python]
obj.period(u, max_time, parameters=None, transient_time=None, tolerance=1e-10, min_period=1, max_period=1000, stability_checks=3)
    \end{lstlisting}
    The arguments \pyinline{u}, \pyinline{parameters}, and \pyinline{transient_time} have been covered already. The argument \pyinline{max_time} is the maximum iteration time. The optional argument \pyinline{tolerance} specifies the numerical tolerance for period detection; it defines the radius of the neighborhood around the initial condition used to determine if the trajectory has returned. The optional arguments \pyinline{min_period} and \pyinline{max_period} set the minimum and maximum periods to consider, respectively. The optional argument \pyinline{stability_checks} determines how many consecutive returns to the neighborhood are required to confirm the period and ensure numerical stability.
    
    For the previous period example, you can proceed as follows:
    \begin{lstlisting}[style=pycon]
>>> u = [1.78, 0.0]  # Initial condition
>>> total_time = 10000  # Total iteration time
>>> transient_time = 5000  # Transient time
>>> k, a, gamma = 8, 0.47, 0.8  # Parameters of the system
>>> parameters = [k, a, gamma]
>>> ds.period(u, total_time, parameters=parameters, transient_time=transient_time)
2
    \end{lstlisting}
    The \pyinline{period} method determines whether a given initial condition leads to a periodic orbit. If so, it returns the period, otherwise, it returns \pyinline{-1} to indicate a quasiperiodic or chaotic orbit. In the last example, the period of the orbit is $2$. By changing continuously the parameters, it is possible to calculate the Lyapunov exponents and the period in the parameter space as well, as shown in Fig.~\ref{fig:fig3}~\cite{RolimSales2024}. The chaotic regions in Fig.~\ref{fig:fig3}(a) are depicted in pink, while the regular regions appear in green. White points indicate locations where $\lambda_1 \rightarrow 0$, corresponds to period-doubling bifurcations. These bifurcations can be more clearly observed in Fig.~\ref{fig:fig3}(b), where colors represent the periodicity of each point. Within the shrimp-shaped domains~\cite{Gallas1993, Gallas1994}, the period progresses from 1 to 2, 4, and so on, eventually leading to chaotic dynamics (marked in white).

\begin{figure}[t]
        \centering
        \includegraphics[width=\linewidth]{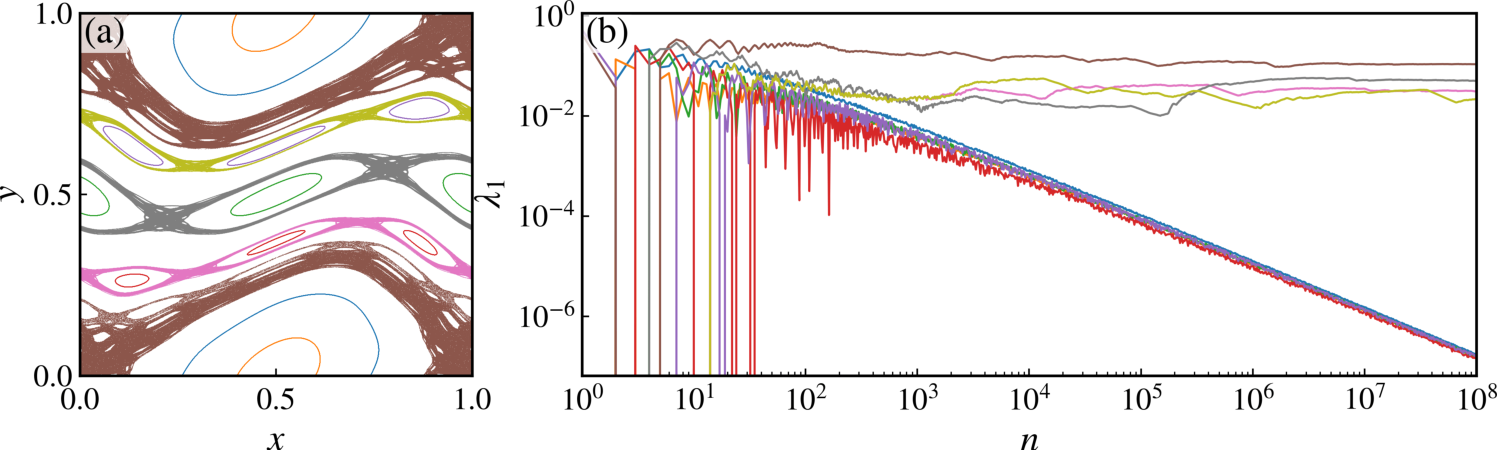}
        \caption{Demonstration of the use of the \pyinline{lyapunov} method of the \pyinline{DiscreteDynamicalSystem} class of \pyinline{pynamicalsys} to compute the time series of the Lyapunov exponents for the standard map [Eq.~\eqref{eq:stdmap}] using specific sample times for the parameter $k = 0.9$.}
        \label{fig:fig4}
    \end{figure}

    In many situations, the history of the Lyapunov exponents is as important as the final value. It tells us how the Lyapunov exponents approach their asymptotic value. To obtain the time evolution of the Lyapunov exponents, set \pyinline{return_history=True} (the default is \pyinline{False}) in the \pyinline{lyapunov} method. To illustrate this feature, let us return to the standard map [Eq.~\eqref{eq:stdmap}]. We choose $k = 0.9$ and select five regular and four chaotic initial conditions [Fig.~\ref{fig:fig4}(a)]. Then, we proceed as follows:
    \begin{lstlisting}[style=pycon]
>>> from pynamicalsys import DiscreteDynamicalSystem as dds
>>> ds = dds(model="standard map")
>>> k = 0.9  # Parameter of the standard map
>>> total_time = 100000000  # Total iteration time
>>> sample_times = np.unique(np.logspace(np.log10(1), np.log10(total_time), 1000).astype(int))  # Sample times to return the LEs
>>> u = np.array([[0.26, 0], [0.4, 0], [0, 0.45], [0.1, 0.25], [0.1, 0.68], [0.06, 0.05], [0, 0.3], [0, 0.6], [0, 0.7]])  # Initial conditions (the first 5 are regular and the last 4 are chaotic)
>>> history_LEs = np.array([ds.lyapunov(u[i], total_time, parameters=k, return_history=True, sample_times=sample_times) for i in range(u.shape[0])])
>>> history_LEs.shape
(9, 836, 2)
    \end{lstlisting}
    This example calculates the history of the Lyapunov exponents for each initial condition at the specified sample times [Fig.~\ref{fig:fig4}(b)]. The largest Lyapunov exponent for the blue, orange, red, green, and purple initial conditions converges toward zero as a power law, indicating regular dynamics. Additionally, they all go to zero at the same rate. The fastest rate of $\lambda_1$ toward zero is obtained exactly on the elliptic points, as Manchein and Beims have demonstrated~\cite{Manchein2013}.
    
    The remaining initial conditions, brown, pink, yellow, and gray, exhibit a positive largest Lyapunov exponent, indicating a chaotic behavior. They seem to converge to a positive value, however, on several occasions, when $\lambda_1$ seems to have converged, its value suddenly decreases and after some time it starts to increase again, and once again, its value decreases. This behavior happens for arbitrarily long times due to the phenomenon of stickiness~\cite{Contopoulos1971, Karney1983, Meiss1983, Chirikov1984, Efthymiopoulos1997, Zaslavsky2002, Zaslavsky2002b, Altmann2006, Cristadoro2008, Contopoulos2008, Contopoulos2010}. The stickiness effect, first identified by Contopoulos~\cite{Contopoulos1971}, influences chaotic orbits as they approach stability islands. When near a regular region, these orbits can become trapped in the vicinity of the islands for arbitrarily long times, exhibiting quasiperiodic-like behavior during this transient phase. As a result, the largest Lyapunov exponent decreases. This phenomenon arises from the complex hierarchical structure of islands-around-islands embedded within the chaotic sea of two-dimensional, area-preserving maps. After escaping one sticky region, a chaotic orbit may later become trapped again in another, repeating the process. Such successive trappings lead to intermittent dynamics, affecting statistical properties, such as diffusion, decay of correlations, and transport.

    Due to the intermittent behavior of chaotic orbits, computing the Lyapunov exponents for longer times might not be the best approach to detect and characterize the stickiness effect. Instead, Szezech et al.~\cite{Szezech2005} proposed the calculation of the finite-time Lyapunov exponents. Before we proceed, we should clarify the term \textit{finite-time}. Strictly speaking, \textit{all} numerical simulations are finite-time. However, their key idea was to compute the Lyapunov exponents not for the whole single trajectory with long times, $N \gg 1$, but rather for shorter time windows, $n \sim 1$, along the same trajectory. Since a chaotic orbit will eventually fill the whole available chaotic component, by considering a total time $N \gg 1$ and calculating the Lyapunov exponent in windows of size $n \ll N$, we obtain a collection of values for the finite-time Lyapunov exponents, $\{\lambda_1^{(i)}\}_{i=1,2,\ldots,M}$, where $M = N / n$, and these values characterize both the intervals where the chaotic orbits are trapped and the intervals where they are in the bulk of the chaotic sea.

    To calculate the finite-time Lyapunov exponents, we use the \pyinline{finite_time_lyapunov} method of the \pyinline{DiscreteDynamicalSystem} class of \pyinline{pynamicalsys}:
    \begin{lstlisting}[style=python]
obj.finite_time_lyapunov(u, total_time, finite_time, parameters=None, method="QR", transient_time=None, log_base=np.e, return_points=False)
    \end{lstlisting}
    Here, \pyinline{total_time} is the total iteration time, $N$, whereas \pyinline{finite_time} is the size of the time windows, $n$, that the Lyapunov exponents are calculated. The optional argument \pyinline{return_points}, if set to \pyinline{True}, tells the \pyinline{finite_time_lyapunov} method to also return the initial condition used to generate the corresponding finite-time Lyapunov exponents. The method thus returns two arrays. Both of them are arrays with $M = N / n$ rows and $d$ columns. Each row contains the finite-time Lyapunov exponents for the first array and for the second array, each row contains the initial condition that generated the respective finite-time Lyapunov exponent.

    The following code snippet illustrates the calculation of the Lyapunov exponents as a function of the parameter $k$ as well as the calculation of the finite-time Lyapunov exponents:
    \begin{lstlisting}[style=pycon]
>>> from pynamicalsys import DiscreteDynamicalSystem as dds
>>> ds = dds(model="standard map")
>>> u = [0.5, 0.25]  # Initial condition for $\lambda$ vs k
>>> k_range = (0, 5, 5000)  # Interval in k
>>> k = np.linspace(*k_range)  # Create the k values
>>> total_time = 5000  # Total iteration time
>>> lyapunov_vs_k = [ds.lyapunov(u, total_time, k[i]) for i in range(k_range[2])]
>>> u = [0.05, 0.05]  # Initial condition for the finite-time Lyapunov exponents
>>> k = 1.5  # Parameter k
>>> total_time = 100000000  # Total iteration time
>>> finite_time = 200  # Finite-time (window size)
>>> ftle, points = ds.finite_time_lyapunov(u, total_time, finite_time, parameters=k, return_points=True)
    \end{lstlisting}

    \begin{figure}[t] 
        \centering
        \includegraphics[width=\linewidth]{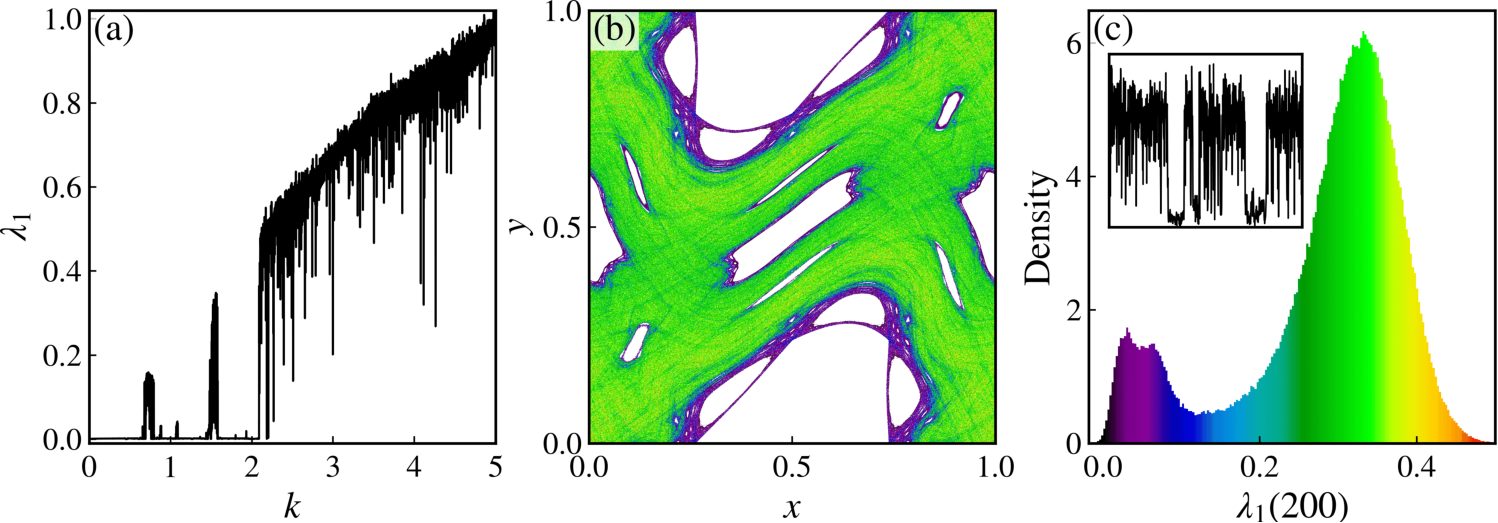}
        \caption{(a) Demonstration of the use of the \pyinline{lyapunov} exponent method of the \pyinline{DiscreteDynamicalSystem} class of \pyinline{pynamicalsys} to calculate the Lyapunov exponents as a function of the parameter $k$ of the standard map [Eq.~\eqref{eq:stdmap}] and (b) and (c) demonstration of the use of the \pyinline{finite_time_lyapunov} exponent method of the \pyinline{DiscreteDynamicalSystem} class of \pyinline{pynamicalsys} to calculate the distribution of finite-time Lyapunov exponents for the standard map with $k = 1.5$.}
        \label{fig:fig5}
    \end{figure}

    In Fig.~\ref{fig:fig5}(a) we show the behavior of $\lambda_1$ as a function of $k$ for the initial condition $(x_0, y_0) = (0.5, 0.25)$. We use the \pyinline{points} output of the \pyinline{finite-time_lyapunov} method as initial conditions and we generate their trajectory for $n = 200$. We color the points of the trajectory according to their respective finite-time Lyapunov exponent value, and this yields Fig.~\ref{fig:fig5}(b). We also calculate the distribution of the finite-time values [Fig.~\ref{fig:fig5}(c)]. We color the histogram according to the finite-time value using the same color code as in Fig.~\ref{fig:fig5}(b). The distribution is a bi-modal distribution, with the largest peak corresponding to the times when the trajectory was on the bulk of the chaotic sea. The smaller peak corresponds to the times when the trajectory was trapped in the vicinity of a stability island. By comparing the colors in Fig.~\ref{fig:fig5}(b) and~\ref{fig:fig5}(c), we can identify the regions in phase space that generated each finite-time value. The smaller values, mainly purple color, correspond to the neighborhood of the islands. The inset in Fig.~\ref{fig:fig5}(c) is the ``times series'' of the finite-time values, i.e., for each time window, we plot the corresponding value of $\lambda_1^{(i)}$. In the inset, we notice mainly two sharp drops in the value of $\lambda_1^{(i)}$. These drops are the times when the trajectory became trapped and they are the reason for the bi-modal distribution.

    \subsection{Linear dependence indexes}
    \label{subsec:LDI}

    As we have discussed in Sec.~\ref{subsec:LE}, given an orthonormal basis $Q \in \mathbb{R}^{d\times d}$, the Jacobian matrix $J$ evolves this basis under the linearized dynamics according to $A_n = J_nQ_{n - 1}$. As we continue to evolve this basis, without performing the QR decomposition, all the basis vectors eventually align with the direction of maximal growth. In light of that, two efficient chaotic indicators have been proposed~\cite{Skokos2003, Skokos2004, Skokos2007}, the second being the generalization of the first. Given a $d$-dimensional discrete-dynamical system $\vb{x}_{n + 1} = \vb{f}(\vb{x}_n)$, where $\vb{f}:\mathbb{R}^d\rightarrow\mathbb{R}^d$, let $J(\vb{x}) = \vb{Df}(\vb{x})$ be the Jacobian matrix of the map $\vb{f}$ at point $\vb{x}$ and $Q \in \mathbb{R}^{d\times 2}$ be a matrix whose columns are two deviation vectors, $\vb{v}_1 \in \mathbb{R}^d$ and $\vb{v}_2\in \mathbb{R}^d$. As the deviation vectors evolve under the linearized dynamics, $\vb{v}_1$ and $\vb{v}_2$ gradually align with the direction of the maximum Lyapunov exponent. The deviation vectors can either align parallel or anti-parallel to the most unstable direction. Thus, we define the parallel alignment index
    \begin{equation}
        \label{eq:pai}
        \mathrm{PAI}(n) = \norm{\frac{\vb{v}_1}{\norm{\vb{v}_1}} - \frac{\vb{v}_2}{\norm{\vb{v}_2}}}
    \end{equation}
    and the antiparallel alignment index
    \begin{equation}
        \label{eq:aai}
        \mathrm{AAI}(n) = \norm{\frac{\vb{v}_1}{\norm{\vb{v}_1}} + \frac{\vb{v}_2}{\norm{\vb{v}_2}}}.
    \end{equation}
    The smallest alignment index (SALI) is then given by
    \begin{equation}
        \label{eq:sali}
        \mathrm{SALI}(n) = \min\qty[\mathrm{PAI}(n), \mathrm{AAI}(n)].
    \end{equation}
    When the two deviation vectors become parallel, $\mathrm{PAI}\rightarrow0$ and $\mathrm{AAI}\rightarrow\sqrt{2}$. On the other hand, when the two deviation vectors become antiparallel, $\mathrm{AAI}\rightarrow0$ and $\mathrm{PAI}\rightarrow\sqrt{2}$. Thus, the SALI captures the information on whether the two deviation vectors tend to have the same direction, either parallel or antiparallel. For $d$-dimensional dynamical systems with $d > 2$, the two vectors tend to become parallel or antiparallel for chaotic orbits~\cite{Voglis1999}, i.e., the SALI tends to zero. On the other hand, for regular orbits, the SALI fluctuates around a positive value. For $2$-dimensional maps, the SALI goes to zero for both chaotic and regular orbits. However, the SALI displays an exponential decay for chaotic orbits whereas for regular orbits, the decay follows a power-law. Thus, by measuring how fast the SALI goes to zero, it is possible to distinguish between regularity and chaos in $2$-dimensional maps as well.

    The generalization of the SALI, the generalized alignment index (GALI), considers the evolution of more than two deviation vectors, i.e., considering the matrix $Q \in \mathbb{R}^{d\times k}$, with $k \leq d$, whose columns are the $k$ deviation vectors $\qty{\vb{v}_i}_{i = 1, 2,\ldots,k}$, the deviation vectors evolve under the linearized dynamics as $Q_n = J_nQ_{n + 1}$. The GALI$_k$ is proportional to the volume elements formed by the $k$ deviation vectors and it is defined as the norm of the wedge product of the $k$ deviation vectors~\cite{Skokos2007}:
    \begin{equation}
        \label{eq:gali}
        \mathrm{GALI}_k = \norm{\vu{v}_1\wedge\vu{v}_2\wedge\ldots\wedge\vu{v}_k},
    \end{equation}
    where $\vu{v} = \vb{v}/\norm{\vb{v}}$ denotes a unit vector. The GALI has been shown to accurately distinguish between regular and chaotic orbits, identify the dimensionality of the space of regular motion, and predict the diffusion of chaotic orbits. Additionally, $\mathrm{GALI}_2 \sim \mathrm{SALI}$. However, GALI$_k$ involves the computation of $\binom{d}{k}$ determinants each time step. For systems with high dimensionality, the computation of GALI thus becomes impractical. For this reason, Antonopoulos and Bountis~\cite{Antonopoulos2006} introduced a new methodology that takes advantage of the linear dependence of the deviation vectors. They realized that both SALI and GALI are, in fact, measures of the linear independence of the deviation vectors. During the time evolution of chaotic orbits, the deviation vectors, which evolve under the linearized dynamics, tend to become linearly dependent over time, i.e., to become asymptotically aligned with each other. Therefore, Antonopoulos and Bountis defined the linear dependence index (LDI) as
    \begin{equation}
        \label{eq:LDI}
        \mathrm{LDI}_k = \prod_{i = 1}^k\sigma_i,
    \end{equation}
    where $\qty{\sigma_i}_{i=1,2,\ldots,k}$ are the singular values of the matrix $Q \in \mathbb{R}^{d\times k}$ whose columns are the $k$ deviation vectors. The singular values are, in fact, a measure of the linear dependence of the deviation vectors. As long as all deviation vectors are linearly independent, $\sigma_i > 0$ for all $i$. As the system evolves in time and the deviation vector aligns with the most unstable direction, $\mathrm{LDI}_k \rightarrow 0$. Thus, at each time step, we compute the singular value decomposition (SVD) of $Q$:
    \begin{equation}
        \label{eq:SVD}
        Q = U\Sigma V^{T},
    \end{equation}
    where $U \in \mathbb{R}^{d\times d}$ is an orthogonal matrix whose columns are the left singular vectors, $\Sigma \in \mathbb{R}^{d\times k}$ is a diagonal matrix containing the non-negative singular values, i.e., $\sigma_i = \Sigma_{ii}$, and $V \in \mathbb{R}^{k\times k}$ is an orthogonal matrix whose columns are the right singular vectors.

    To illustrate the computation of the LDI's, we consider a four-dimensional symplectic map, given by
    \begin{equation}
        \label{eq:4dsym}
        \begin{aligned}
            x_{n+1}^{(1)} &= x_{n}^{(1)} + x_{n}^{(2)}\bmod{2\pi},\\
            x_{n+1}^{(2)} &= x_{n}^{(2)} - \epsilon_1\sin\qty(x_{n}^{(1)} + x_{n}^{(2)}) - \xi\qty[1 - \cos(x_{n}^{(1)} + x_{n}^{(2)} + x_{n}^{(3)} + x_{n}^{(4)})] \bmod{2\pi},\\
            x_{n+1}^{(3)} &= x_{n}^{(3)} + x_{n}^{(4)} \bmod{2\pi},\\
            x_{n+1}^{(4)} &= x_{n}^{(4)} - \epsilon_2\sin\qty(x_{n}^{(3)} + x_{n}^{(4)}) - \xi\qty[1 - \cos(x_{n}^{(1)} + x_{n}^{(2)} + x_{n}^{(3)} + x_{n}^{(4)})] \bmod{2\pi}.
        \end{aligned}
    \end{equation}
    This map is composed of two coupled standard maps with parameters $\epsilon_1$ and $\epsilon_2$. The parameter $\xi$ is the coupling strength. For $\xi = 0$, the two maps behave independently. We set the parameters to $(\epsilon_1, \epsilon_2, \xi) = (0.5, 0.1, 0.001)$ and we consider two initial conditions. One regular: $(x_0^{(1)}, x_0^{(2)}, x_0^{(3)}, x_0^{(4)}) = (0.5, 0, 0.5, 0)$ [red curve in Fig.~\ref{fig:fig5}(a)] and one chaotic: $(x_0^{(1)}, x_0^{(2)}, x_0^{(3)}, x_0^{(4)}) = (3.0, 0, 0.5, 0)$ [blue curve in Fig.~\ref{fig:fig5}(a)]. 

\begin{figure}[t]
        \centering
        \includegraphics[width=\linewidth]{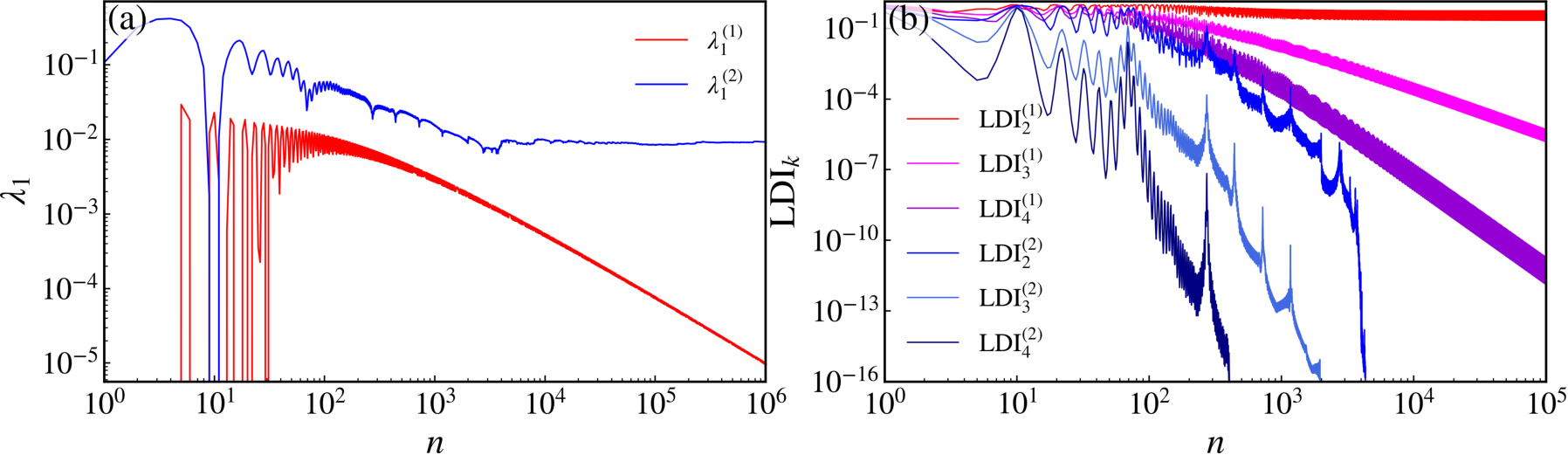}
        \caption{Demonstration of the use of the \pyinline{LDI} method of the \pyinline{DiscreteDynamicalSystem} class of \pyinline{pynamicalsys} for two different initial conditions of a four-dimensional symplectic map [Eq.~\eqref{eq:4dsym}].}
        \label{fig:fig6}
    \end{figure}

    To calculate the LDI's, we use the \pyinline{LDI} method of the \pyinline{DiscreteDynamicalSystem} class from \pyinline{pynamicalsys}:
    \begin{lstlisting}[style=python]
obj.LDI(u, total_time, k, parameters=None, return_history=False, sample_times=None, tol=1e-16, transient_time=None, seed=13)
    \end{lstlisting}
    Here, the integer \pyinline{k} specifies the number of deviation vectors. Optional arguments include \pyinline{return_history}, which determines whether to return the full-time evolution of the LDI's (default is \pyinline{False}), \pyinline{sample_times}, a list of times at which to sample the LDI (default is \pyinline{None}), \pyinline{tol}, the numerical tolerance for stopping the calculation (default is $10^{-16}$), \pyinline{transient_time}, which allows skipping an initial transient phase (default is \pyinline{None}), and \pyinline{seed}, which sets the random seed to ensure reproducibility when generating the initial deviation vectors (default is \pyinline{13}). The following code illustrates how to calculate the $\mathrm{LDI}_k$ for different initial conditions [Fig.~\ref{fig:fig6}(b)]:
    \begin{lstlisting}[style=pycon]
>>> from pynamicalsys import DiscreteDynamicalSystem as dds
>>> ds = dds(model="4D symplectic map")        
>>> info = ds.info
>>> info["parameters"]  # Get the info about the order of the parameters
['eps1', 'eps2', 'xi']
>>> u = np.array([[0.5, 0.0, 0.5, 0.0], [3.0, 0.0, 0.5, 0.0]])  # Initial conditions
>>> parameters = [0.5, 0.1, 0.001]  # Define the parameters
>>> k = [2, 3, 4]  # The numbers of deviation vectors
>>> total_time = int(1e5)  # Total iteration time
>>> ldi = np.zeros((u.shape[0], total_time, len(k)))
>>> for i in range(len(k)): 
...    for j in range(u.shape[0]):
...        ldi[j, :, i] = ds.LDI(u[j], total_time, k[i], parameters=parameters, return_history=True)
    \end{lstlisting}

    Figure~\ref{fig:fig6} shows the largest Lyapunov exponent and the LDI's for $k = 2$, $k = 3$, and $k = 4$, for two initial conditions, one regular and one chaotic, of the four-dimensional symplectic map [Eq.~\eqref{eq:4dsym}]. The behavior of the LDI's for the regular orbit (red, pink, and purple) and the chaotic orbit (blue, light blue, and dark blue) differs fundamentally. For instance, LDI$_2$ for the regular orbit (red curve) oscillates around a positive value and does not converge to zero. On the other hand, the LDI$_2$ for the chaotic orbit (blue curve) decreases toward zero exponentially fast, reaching the machine precision of $10^{-16}$ for less than $4\times10^3$ iterations. The LDI$_3$ and LDI$_4$ for the chaotic orbits (light blue and dark blue curves, respectively) decreases to zero even faster. And even though the LDI$_3$ and LDI$_4$ for the regular orbit (pink and purple curves, respectively) also decrease to zero, they do so in a power law, which is much slower than the exponential decay of the chaotic orbit. Therefore, the LDI is an extremely fast and accurate chaotic indicator.

    Antonopoulos and Bountis~\cite{Antonopoulos2006} demonstrated that $\mathrm{LDI}_k \sim \mathrm{GALI}_k$ and thus $\mathrm{LDI}_2 \sim \mathrm{SALI}$. The red, fuchsia, and purple curves in Fig.~\ref{fig:fig6}(b) are the LDI curves for $k = 2$, $k = 3$, and $k = 4$, respectively. These curves correspond to the regular initial condition. We observe that $\mathrm{LDI}_2$ does not tend to zero, but it rather oscillates around a positive value, corroborating the statement of Antonopoulos and Bountis. The LDI$_3$ and LDI$_4$, on the other hand, decay to zero but following a power-law~\cite{Skokos2007}. The LDI for the chaotic initial condition [shades of blue in Fig.~\ref{fig:fig6}(b)] all decay to zero exponentially, analogously to GALI~\cite{Skokos2007}.

    Therefore, computing either GALI$_k$ or LDI$_k$ yields the same result with the difference of the LDI demanding less CPU time. Additionally, there is no real difference between computing SALI or LDI$_2$. However, in this case, SALI is computationally faster than the LDI$_2$. Thus, if one is studying a two-dimensional map or for some reason is only interested in two deviation vectors, a faster alternative is to compute the SALI directly instead of the LDI$_2$:
    \begin{lstlisting}[style=python]
obj.SALI(u, total_time, parameters=None, return_history=False, sample_times=None, tol=1e-16, transient_time=None, seed=13)
    \end{lstlisting}
    It is about 9 times computationally faster than the \pyinline{LDI} method, which makes sense since to calculate the SALI, we only need to evaluate the norms of the deviation vectors while the LDI's require the singular value decomposition at each time step.

    \subsection{Weighted Birkhoff averages}

    The weighted Birkhoff average~\cite{Das2016, Das2017, Das2018, Sander2020, Meiss2021, Duignan2022} is a powerful tool to classify the dynamics as regular or chaotic in a Hamiltonian system, without the problems introduced by the self-similar hierarchical structure, in which islands and chaotic orbits are mixed together in arbitrarily fine scales~\cite{Meiss2021, Sander2020}. For a mapping of the form ${\bf v}_{n + 1} = {\bf M}^n({\bf v}_0)$. The Birkhoff average of some function $f({\bf v})$ along this trajectory in phase space is defined as
    \begin{equation}
        \label{eq:ba}
        B_N(f)({\bf v}_0) = \frac{1}{N} \sum_{n=0}^{N-1} f \circ {\bf M}^n({\bf v}_0).
    \end{equation}
        
    The Birkhoff ergodic theorem~\cite{cornfeld2012ergodic} states that time averages of the function $f$ along the trajectory converge to the phase space averages as $N \rightarrow \infty$
    \begin{equation}
        \label{eq:bet}
        \frac{1}{N} \sum_{n=0}^{N-1} f \circ {\bf M}^n({\bf v}_0) \rightarrow \int f d\mu,
    \end{equation}
    where $\mu$ is an invariant probability measure. However, convergence can be slow—scaling as $N^{-1}$ for quasiperiodic orbits and $N^{-1/2}$ for chaotic ones—due to edge effects from finite-time segments.

    To mitigate this, a weighted Birkhoff average is introduced:
    \begin{equation}
        \label{eq:wba}
        W\!B_N(f)({\bf v}_0) = \sum_{n=0}^{N-1} w_{n,N} f \circ {\bf M}^n({\bf v}_0),
    \end{equation}
    where 
    \begin{equation}
        \label{eq:weights}
        w_{n,N} = \frac{g(n/N)}{\sum_{n=0}^{N-1} g(n/N)},
    \end{equation}
    with an exponential bump function
    \begin{equation}
        \label{eq:bump}
        g(z) = \begin{cases}
              \exp\{-{\lbrack z(1-z)\rbrack}^{-1} \}, & \text{if $0 < z < 1$}, \\
              0, & \text{otherwise}.
           \end{cases}
    \end{equation}
    This choice ensures smooth vanishing at the endpoints and preserves regularity. When $f$, $g$, and ${\bf M}$ are $C^\infty$, the convergence becomes super-polynomial~\cite{Das2018}:
    \begin{equation}
        \label{eq:super}
        \left\vert W\!B_N(f)({\bf v}_0) - \int f d\mu \right\vert \le C_m N^{-m}.
    \end{equation}
    Notably, this acceleration applies only to regular orbits. Moreover, the convergence rate is largely independent of the choice of $f$, allowing the use of simple observables, such as $f(x, y) = \sin(x + y)$~\cite{Das2016} or $f(x) = \cos x$~\cite{Sander2020}.

    To distinguish between regular and chaotic dynamics, one computes $2N$ iterations of ${\bf M}$ and compares:
    \begin{equation}
        \label{eq:dig}
        {\mathrm{dig}} = - \log_{10} \left\vert W\!B_N(f)({\bf v}_0) - W\!B_N(f)({\bf v}_{N+1}) \right\vert.
    \end{equation}
       
    A high value of $\mathrm{dig}$ indicates fast convergence and thus regular motion. Low values suggest chaos, though comparisons between chaotic orbits are not meaningful in the Lyapunov sense.

\begin{figure}[t]
        \centering
        \includegraphics[width=\linewidth]{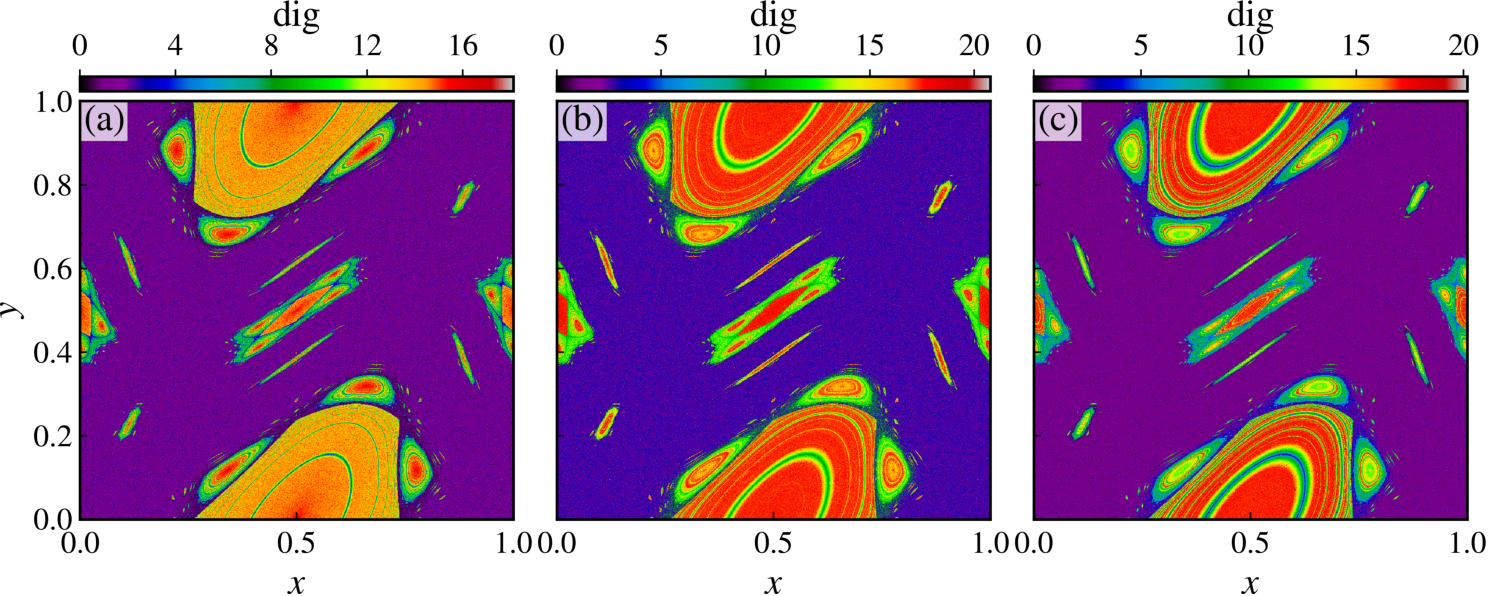}
        \caption{Demonstration of the use of the \pyinline{dig} method of the \pyinline{DiscreteDynamicalSystem} class of \pyinline{pynamicalsys} to calculate dig for a grid of uniformly distributed initial conditions for the standard map [Eq.~\eqref{eq:stdmap}] with $k = 1.5$ and for the functions (a) $f(x) = \cos\qty(2\pi x)$, (b) $f(x) = \sin\qty(2 \pi x)$, and (c) $f(x, y) = \sin\qty[2\pi (x + y)]$.}
        \label{fig:fig7}
    \end{figure}

    To calculate dig, we use the \pyinline{dig} method of the \pyinline{DiscreteDynamicalSystem} class from \pyinline{pynamicalsys}:
    \begin{lstlisting}[style=python]
obj.dig(u, total_time, parameters=None, func=lambda x: np.cos(2 * np.pi * x[:, 0], transient_time=None)
    \end{lstlisting}
    Here, \pyinline{func} corresponds to the function $f$ discussed above. By default, it uses $f(x) = \cos(2\pi x)$ [Fig.~\ref{fig:fig7}(a)]. To use a different function, let us say $f(x) = \sin(2 \pi x)$ [Fig.~\ref{fig:fig7}(b)] or $f(x, y) = \sin(2 \pi (x + y))$ [Fig.~\ref{fig:fig7}(c)], we can pass a different lambda function to the \pyinline{dig} method: \pyinline{func}=lambda x: sin(2 * np.pi * x[:, 0]) or \pyinline{func}=lambda x: sin(2 * np.pi * (x[:, 0] + y[:, 0])), for instance. The following code snippet illustrates the calculation of dig using these three functions $f$:
    \begin{lstlisting}[style=pycon]
>>> from pynamicalsys import DiscreteDynamicalSystem as dds
>>> ds = dds(model="standard map")
>>> grid_size = 1000  # Size of the grid in phase space
>>> x = np.linspace(0, 1, grid_size)  # Create the grid
>>> y = np.linspace(0, 1, grid_size)
>>> X, Y = np.meshgrid(x, y)
>>> u = np.array([X.flatten(), Y.flatten()]).T
>>> k = 1.5  # Parameter of the map
>>> total_time = 10000  # Total iteration time
>>> dig = [ds.dig(u[i], total_time, parameters=k) for i in range(u.shape[0])]
>>> function_b = lambda x: np.sin(2 * np.pi * x[:, 0])
>>> dig2 = [ds.dig(u[i], total_time, parameters=k, func=function_b) for i in range(u.shape[0])]
>>> function_c = lambda x: np.sin(2 * np.pi * (x[:, 0] + x[:, 1]))
>>> dig3 = [ds.dig(u[i], total_time, parameters=k, func=function_c) for i in range(u.shape[0])]
    \end{lstlisting}

    All functions $f$ are able to distinguish the regular and chaotic regions, the choice of this function is rather arbitrary, as long as it is sufficiently smooth, $C^\infty$, and maps into a finite-dimensional real vector space~\cite{Das2018}. As expected, the quasiperiodic orbits exhibit a high value of dig, while chaotic orbits have a small value of dig. Since the weighted Birkhoff average does not improve the convergency of a chaotic orbit, we cannot compare two values of dig for two different chaotic orbits. For instance, we cannot say that an orbit with $\mathrm{dig}=1$ is ``more chaotic'' than one with $\mathrm{dig}=2$. Nevertheless, the weighted Birkhoff average can efficiently distinguish regular and chaotic regions, being simple and faster to compute. It helps the calculation of the dimension of the boundary of the chaotic sea and the islands~\cite{Sales2022}, the classification of regions in a drift $\mathbf{E}\times\mathbf{B}$ model~\cite{Souza2023}, and can also be extended to flows~\cite{Duignan2023}.
        
    \subsection{Recurrence time entropy}

    The recurrence plot (RP) was introduced in 1987 by Eckmann \textit{et al.}~\cite{Eckmann1987} as a graphical representation of the recurrences of time series of dynamical systems in its $d$-dimensional phase space. For a given trajectory $\vb{x}_i \in \mathbb{R}^d$ ($i = 1, 2, \ldots, N$) of length $N$, the $N \times N$ recurrence matrix is defined as
    \begin{equation}
        R_{ij} = H\qty(\epsilon - \|\vb{x}_i - \vb{x}_j\|),
    \end{equation}
    where $H(\cdot)$ is the Heaviside unit step function, $\epsilon$ is a small threshold, and $\|\vb{x}_i - \vb{x}_j\|$ is the distance between states $\vb{x}_i$ and $\vb{x}_j$ in phase space measured in terms of a suitable norm. The most commonly used norms are the Euclidean norm and the maximum (or supremum) norm, defined as
    \begin{equation}
        \begin{aligned}
            \norm{\vb{x}}_2 &= \qty(\sum_{i = 1}^d \abs{x_i}^2)^{1/2},\\
            \norm{\vb{x}}_\infty &= \max_{i}\qty(\abs{x_i}),
        \end{aligned}
    \end{equation}
    respectively. Both of these norms yield similar results. However, the maximum norm is computationally faster and it results in more recurrent points for a fixed threshold $\epsilon$~\cite{Marwan2007}. Therefore, the maximum norm is more commonly used.

    The recurrent states are represented by the value $1$ in the symmetric, binary recurrence matrix $\vb{R}$, whereas the nonrecurrent ones are represented by the value $0$. Since it is numerically impossible to find \emph{exactly} recurrent states, i.e., $\vb{x}_i = \vb{x}_j$, two states are said to be recurrent if they are sufficiently close to each other up to a distance $\epsilon$. The distance $\epsilon$ has to be carefully chosen. If $\epsilon$ is set too large, nearly every state is recurrent with every other state. On the other hand, if $\epsilon$ is chosen too small, there will be few recurrent states. Hence, a compromise has to be made between choosing $\epsilon$ as small as possible while resulting in a sufficient number of recurrent states. There is no general rule on choosing $\epsilon$. However, a few options have been proposed in the literature and each one of them has its own advantages and disadvantages depending on the purpose of the study. For instance, an alternative is to choose $\epsilon$ such that the recurrence point density, i.e., the recurrence rate, is fixed~\cite{Zbilut2002, Kraemer2018}. While this eliminates the issue of obtaining few recurrent states, it only shifts the problem to finding the optimal value of the recurrence rate. Another possibility is to define $\epsilon$ as a fraction of the time series standard deviation~\cite{Thiel2002, Marwan2007, Schinkel2008}. This has been proved efficient when distinguishing between dynamical regimes and analyzing dynamical transition~\cite{Ngamga2007, Ngamga2008, Sales2023, Souza2024}. 
    
    Graphically, the recurrent states are represented by a colored dot and the recurrence matrix exhibits different patterns depending on the dynamics of the underlying system. These patterns are composed of mainly four distinct structures. They are (i) isolated recurrence points, (ii) diagonal lines, (iii) vertical lines, and (iv) white (non-recurrent) vertical lines. The recurrence matrix can be calculated using the \pyinline{recurrence_matrix} method of the \pyinline{DiscreteDynamicalSystem} class from \pyinline{pynamicalsys}:
    \begin{lstlisting}[style=python]
obj.recurrence_matrix(u, total_time, parameters=None, transient_time=None, **kwargs)
    \end{lstlisting}
    This method, given an initial condition $\vb{u} \in \mathbb{R}^d$, stored in the array \pyinline{u}, returns the recurrence matrix $\vb{R} \in \mathbb{R}^{d\times d}$. The optional arguments include \pyinline{metric}, which defines which norm to use. By default, the \pyinline{recurrence_matrix} method uses the supremum norm (\pyinline{metric="supremum"}). The method also supports the Euclidean norm \pyinline{metric="euclidean"} and the Manhattan norm ($L_1$ norm) \pyinline{metric="manhattan"}. The threshold $\epsilon$ setting is done via the \pyinline{threshold} optional argument. By default, it is set to \pyinline{threshold=0.1}. Additionally, the optional argument \pyinline{threshold_std} determines whether to use the threshold in units of the standard deviation of the trajectory generated by the initial condition \pyinline{u}. By default, it is set to \pyinline{threshold_std=True}. In this case, \pyinline{threshold=0.1} means $10\%$ of the trajectory's standard deviation. Regarding the standard deviation, for a one-dimensional system, it is simply the standard deviation of the whole trajectory. However, for higher-dimensional systems, we define a standard deviation vector where each component contains the standard deviation of the corresponding component of the trajectory:
    \begin{equation}
        \boldsymbol{\sigma} = (\sigma_1, \sigma_2,\ldots,\sigma_d)^T,
    \end{equation}
    where $\sigma_i = \mathrm{std}(\vb{x}_i)$. Then, we define the standard deviation of the trajectory as the norm of the standard deviation vector $\boldsymbol{\sigma}$ (supremum, Euclidean, or Manhattan) using the optional argument \pyinline{std_metric} (default is \pyinline{std_metric="supremum"}).

\begin{figure}[t]
        \centering
        \includegraphics[width=\linewidth]{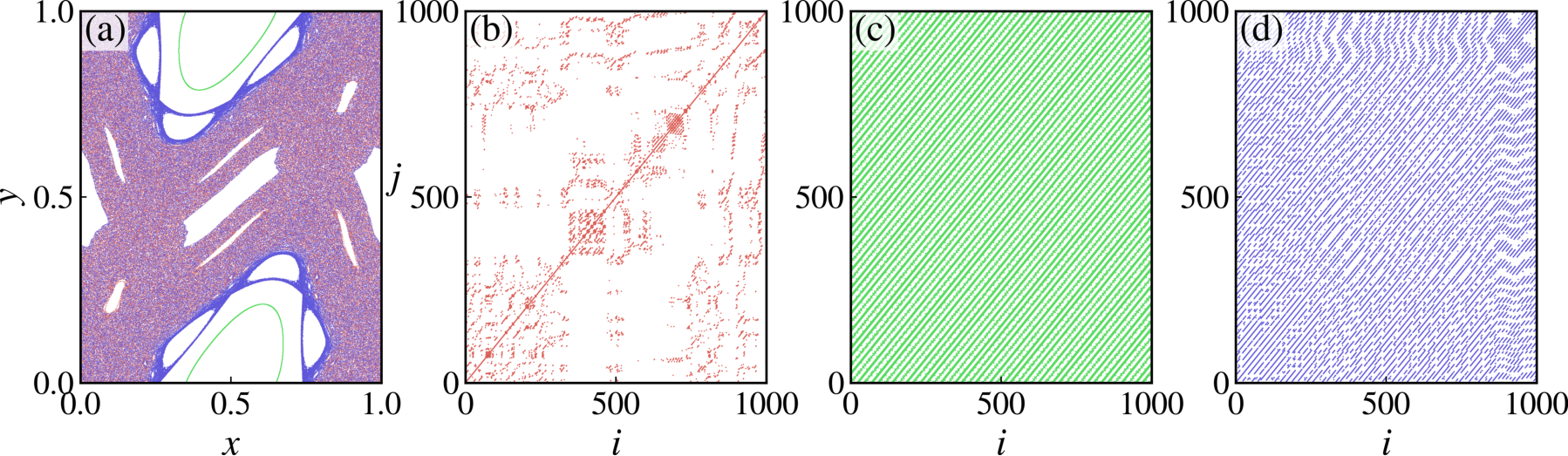}
        \caption{Demonstration of the use of the \pyinline{recurrence_matrix} method of the \pyinline{DiscreteDynamicalSystem} class of \pyinline{pynamicalsys} for three different initial conditions for the standard map [Eq.~\eqref{eq:stdmap}] with $k = 1.5$.}
        \label{fig:fig8}
    \end{figure}

    The following code calculates the recurrence matrix for three different initial conditions for the standard map [Eq.~\eqref{eq:stdmap}] using 10\% of the trajectories' standard deviation as the threshold [Figs.~\ref{fig:fig8}(b)--\ref{fig:fig8}(d)]:
    \begin{lstlisting}[style=pycon]
>>> from pynamicalsys import DiscreteDynamicalSystem as dds
>>> ds = dds(model="standard map")
>>> u = [[0.05, 0.05], [0.35, 0.0], [0.42, 0.2]]  # Initial conditions
>>> k = 1.5  # Parameter of the map
>>> total_time = 1000  # Total iteration time
>>> recmats = [ds.recurrence_matrix(u[i], total_time, parameters=k) for i in range(len(u))]
    \end{lstlisting}

    Figure~\ref{fig:fig8}(a) shows the trajectories for three distinct initial conditions, namely, (red) $(x, y) = (0.05, 0.05)$, (green) $(x, y) = (0.35, 0.0)$, and (blue) $(x, y) = (0.42, 0.2)$. The first one is an initial condition in the bulk of the chaotic sea and the second one is inside of an island. The third one, however, is an initial condition located at the sticky region around the period-6 islands. For the first iterations, this initial condition remains trapped inside the sticky region, hence a high density of blue points around the period-6 islands. Figures~\ref{fig:fig8}(b)-\ref{fig:fig8}(d) show the recurrence matrices for the initial conditions in Fig.~\ref{fig:fig8}(a). The recurrence matrix for the chaotic orbit [Fig.~\ref{fig:fig8}(b)] exhibits few diagonal lines while the recurrence matrix for the quasiperiodic orbit [Fig.~\ref{fig:fig8}(c)] is composed mainly of diagonal lines. The recurrence matrix for the sticky orbit [Fig.~\ref{fig:fig8}(d)], on the other hand, is not as regular as the quasiperiodic one but is more regular than the chaotic one. The recurrence matrix can also be calculated using only a given time series with the \pyinline{recurrence_matrix} method of the \pyinline{TimeSeriesMetrics} of \pyinline{pynamicalsys} (see the documentation). 

    Therefore, the RPs of different dynamical processes are qualitatively different. In order to quantify the structures in RPs, several measures based on the diagonal and vertical lines have been proposed, such as the determinism and the laminarity, for instance. For a complete discussion on these and other measures, we refer the reader to Refs.~\cite{Webber1994, Gao1999, Shockley2002, Marwan2002, Marwan2002b, Marwan2007, Marwan2008} and references therein. Measures based on the white vertical lines, i.e., the vertical distance between two diagonal lines, have also been proposed~\cite{Ngamga2007, Ngamga2008, Kraemer2018, Sales2023}. The white vertical lines of an RP are an estimate of the return times of the corresponding trajectory~\cite{Gao2000, Zou2007, Ngamga2012}, which is the time an orbit takes to return to the neighborhood of a previous point on the orbit. The RP for the chaotic orbit [Fig.~\ref{fig:fig8}(b)] shows no regularity in the vertical distances between the diagonal lines while the RP of the quasiperiodic process [Fig.~\ref{fig:fig8}(c)] consists diagonal lines with different distances between them. The vertical distances in the RP of the sticky orbit [Fig.~\ref{fig:fig8}(d)], on the other hand, are more regular than the chaotic case but it is not as regular as the quasiperiodic case. Thus, the RP of the sticky orbit has an intermediate complexity when compared to the RP of the quasiperiodic and chaotic ones.

    Now, quasiperiodic processes yield three return times. This fact is stated by Slater's theorem~\cite{Slater1950, Slater1967, Mayer1988}: for any irrational rotation, with rotation number $\omega$, over a unit circle, there are at most three different return times to a connected interval of size $\delta < 1$. Additionally, the third return time is the sum of the other two, and two of these three return times are consecutive denominators in the continued fraction expansion of $\omega$. Slater's theorem has been applied to study two-dimensional, area-preserving mappings~\cite{Zou2007b, Baroni2019} as well as to study the parameter space of a one-dimensional mapping~\cite{Mugnaine2022}. It has also been employed to locate the position in phase space of invariant curves in Hamiltonian systems as well as to find the critical parameter values at which these curves break up~\cite{Abud2015, Hermes2022, Huggler2022}.
    
    Therefore, due to the intrinsic property of dynamical systems that quasiperiodic dynamics result in three recurrence times, measures based on the distribution of recurrence times are excellent alternatives to the characterization of the dynamics. We introduce, then, the recurrence time entropy (RTE), i.e., the Shannon entropy of the distribution of white vertical lines (recurrence times), estimated from the RP. Formally, the total number of white vertical lines of length $\ell$ is given by the histogram
    \begin{equation}
        \label{eq:histogram}
        P(\ell) = \sum_{i, j = 1}^N R_{i, j - 1}R_{i, j + \ell}\prod_{k = 0}^{\ell - 1}(1 - R_{i, j + k}),
    \end{equation}
    such that the RTE is defined as~\cite{Kraemer2018}
    \begin{equation}
        \label{eq:rte}
        \mathrm{RTE} = -\sum_{\ell = \ell_{\text{min}}}^{\ell_\text{max}}p(\ell)\log p(\ell),
    \end{equation}
    where $\ell_{\text{min}}$ ($\ell_{\text{max}}$) is the length of the shortest (longest) white vertical line, $p(\ell) = P(\ell) / \mathcal{N}$ is the relative distribution of white vertical lines of length $\ell$ and $\mathcal{N}$ is the total number of white vertical line segments. The evaluation of the histogram given by Eq.~\eqref{eq:histogram} should be done carefully. Due to the finite size of an RP, the distribution of white vertical lines might be biased by the border lines, i.e., the lines that begin and end at the borders of an RP. These lines are cut short by the borders of the RP and their length is measured incorrectly. This influences measures such as the RTE~\cite{Kraemer2019}. To avoid such border effects, we exclude from the distribution the white vertical lines that begin and end at the border of the RP.

    Originally, the RTE was introduced with no connections to RPs~\cite{Little2007}, and it has been shown that it provides a good estimate for the Kolmogorov-Sinai entropy~\cite{Baptista2010}. The RTE has also been successfully applied to detect sticky regions in two-dimensional, area-preserving mappings~\cite{Sales2023, Souza2024} and to detect dynamical transitions in a fractional cancer model~\cite{Gabrick2023}. A periodic orbit, which has only one return time, yields $\mathrm{RTE} = 0$. A quasiperiodic orbit, which has three return times, is characterized by a low value of RTE, whereas a chaotic orbit leads to a high value of RTE. Since the RP of a sticky orbit has an intermediate complexity, the RTE for such an orbit is smaller than the chaotic case but larger than the quasiperiodic case.

    The RTE is the only recurrence-based measure implemented in \pyinline{pynamicalsys}. That is not the aim of the package. We have chosen to implement the RTE due to its ability to detect different hierarchical levels in the islands-around-islands structure in two-dimensional, area-preserving maps and for being able to detect dynamical transitions~\cite{Sales2023, Gabrick2023, Souza2024}. For a complete package on RQAs, we refer the reader to the \pyinline{pyunicorn} package~\cite{pyunicorn}. 
    
    To calculate the RTE, we use the \pyinline{recurrence_time_entropy} method of the \pyinline{DiscreteDynamicalSystem} class of \pyinline{pynamicalsys}:
    \begin{lstlisting}[style=python]
obj.recurrence_time_entropy(u, total_time, parameters=None, transient_time=None, **kwargs)        
    \end{lstlisting}
    It returns the RTE for the given initial conditions. The optional arguments here include the \pyinline{metric}, \pyinline{threshold}, \pyinline{threshold_std}, and \pyinline{std_metric} we have discussed already. It also includes the \pyinline{lmin}, which corresponds to the $\ell_{\mathrm{min}}$ value in Eq.~\eqref{eq:rte} (default is \pyinline{lmin=1}). The \pyinline{recurrence_time_entropy} method can also return the final state (set \pyinline{return_final_state=True}), the recurrence matrix used in the RTE calculation (set \pyinline{return_recmat=True}), and the distribution of white vertical lines $p(\ell)$ (set \pyinline{return_p=True}).

\begin{figure}[t]
        \centering
        \includegraphics[width=\linewidth]{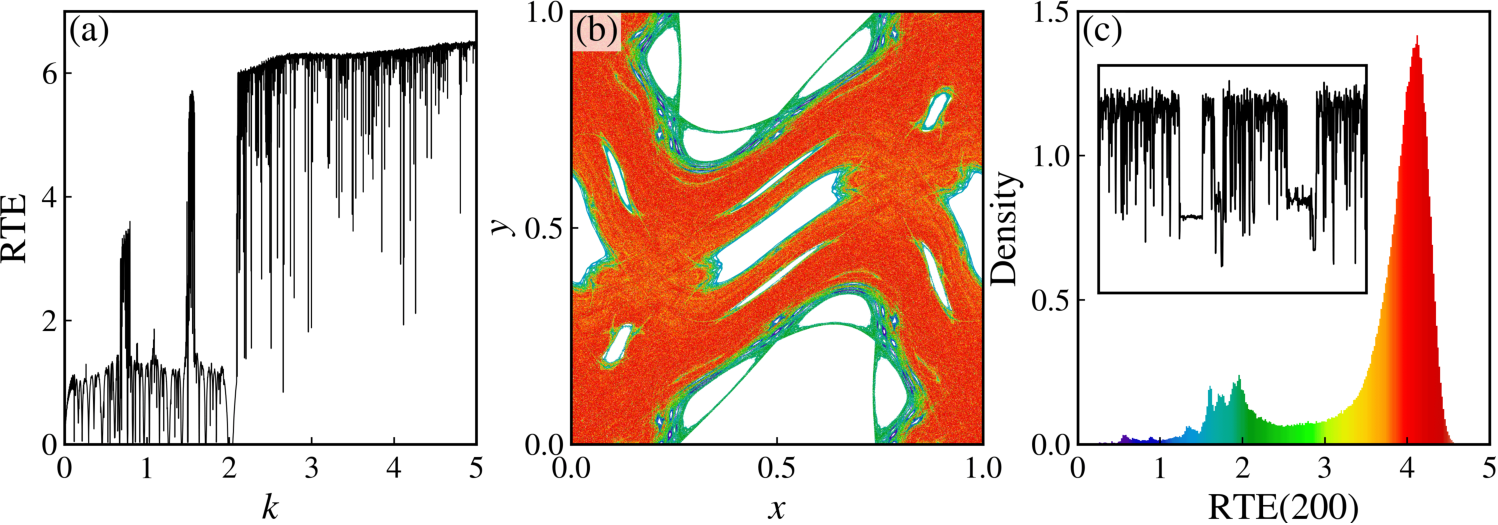}
        \caption{(a) Demonstration of the use of the \pyinline{recurrence_time_entropy} method of the \pyinline{DiscreteDynamicalSystem} class of \pyinline{pynamicalsys} to calculate the RTE as a function of the parameter $k$ of the standard map [Eq.~\eqref{eq:stdmap}] and (b) and (c) demonstration of the use the \pyinline{finite_time_recurrence_time_entropy} method of the \pyinline{DiscreteDynamicalSystem} class of \pyinline{pynamicalsys} to calculate the distribution of finite-time RTE for the standard map with $k = 1.5$.}
        \label{fig:fig9}
    \end{figure}

    The following code demonstrates how to calculate the RTE for the standard map [Eq.~\eqref{eq:stdmap}] as a function of the nonlinearity parameter $k$ [Fig.~\ref{fig:fig9}(a)]:
    \begin{lstlisting}[style=pycon]
>>> from pynamicalsys import DiscreteDynamicalSystem as dds
>>> ds = dds(model="standard map")
>>> u = [0.5, 0.25]  # Initial condition
>>> k_range = (0, 5, 5000)  # Interval in k
>>> k = np.linspace(*k_range)  # Create the k values
>>> total_time = 5000  # Total iteration time
>>> rte = [ds.recurrence_time_entropy(u, total_time, parameters=k[i]) for i in range(k_range[2])]
    \end{lstlisting}

    Figure~\ref{fig:fig9}(a) shows that the RTE distinguishes chaos and regularity successfully [cf.~Fig.~\ref{fig:fig5}(a)]. And similarly to the Lyapunov exponents, the RTE can also be used to detect sticky orbits~\cite{Sales2023, Souza2024} by calculating the finite-time RTE, i.e., for a long trajectory of length $N$ (\pyinline{total_time}), we calculate the RTE in windows of size $n \ll N$ (\pyinline{finite_time}). We use the \pyinline{finite_time_recurrence_time_entropy} method of the \pyinline{DiscreteDynamicalSystem} class of \pyinline{pynamicalsys}:
    \begin{lstlisting}[style=python]
>>> obj.finite_time_recurrence_time_entropy(u, total_time, finite_time, parameter=None, return_points=False, **kwargs)
    \end{lstlisting}
    It returns an array of size $M = N / n$ with the value of the finite-time RTE for each time window. The optional arguments here include the \pyinline{metric}, \pyinline{threshold}, \pyinline{threshold_std}, and \pyinline{std_metric} and also the \pyinline{return_points}. When set to \pyinline{return_points=True}, the method also returns the initial conditions that generate the corresponding finite-time RTE value. The following code snippet illustrates the use of the \pyinline{finite_time_recurrence_entropy} method to calculate the finite-time RTE distribution [Fig.~\ref{fig:fig9}(b) and~\ref{fig:fig9}(c)]:
    \begin{lstlisting}[style=pycon]
>>> from pynamicalsys import DiscreteDynamicalSystem as dds
>>> ds = dds(model="standard map")
>>> u = [0.05, 0.05]  # Initial condition
>>> k = 1.5  # Parameter of the map
>>> total_time = 100000000  # Total iteration time
>>> finite_time = 200  # Finite time
>>> ftrte, points = ds.finite_time_recurrence_time_entropy(u, total_time, finite_time, parameters=k, return_points=True)
    \end{lstlisting}

    The distribution of the finite-time RTE is similar to the distribution of the finite-time Lyapunov exponent [Fig.~\ref{fig:fig5}(c)]. With the RTE, however, it is possible to detect more than one mode, i.e., the smaller mode in Fig.~\ref{fig:fig5}(c) is, in fact, composed of several smaller modes, as suggested by Harle and Feudel~\cite{Harle2007HierarchyFTLEs}. By inspecting the phase space positions that generate the smaller peaks in the distribution, we notice that smaller values of RTE are associated with inner levels in the hierarchical structure of islands-around-islands~\cite{Sales2023, Souza2024}. The inset in Fig.~\ref{fig:fig9}(c) corresponds to the same windows of the inset in Fig.~\ref{fig:fig5}(c). We notice the sharp drops in the value of the finite-time RTE, which leads to the multi-modal distribution we see in Fig.~\ref{fig:fig9}(c).

    \subsection{Hurst exponent}

    The Hurst exponent was introduced by H. E. Hurst in 1951 to model the cyclical patterns of the Nile floods~\cite{Hurst1951} and is a key metric for assessing long-term memory in time series, revealing the extent to which data points are persistently or anti-persistently correlated over time. Several algorithms have been proposed to numerically estimate the Hurst exponent~\cite{Geweke1983, Peng1994, Alessio2002, Zhang2024}, however, the rescaled range analysis (R/S analysis)~\cite{Hurst1951, Mandelbrot1968, Mandelbrot1969} has become the standard approach to its estimation. The approach is as follows: Given an $1$-dimensional time series, $\vb{x} = (x_1, x_2, \ldots, N)$ of length $N$, we divide the time series into $\kappa$ non-overlapping subseries of length $\ell$, $\qty{\vb{P}_k(\ell)}_{k=1,2,\ldots,\kappa}$, such that $\kappa = N / \ell$. For each subseries $\vb{P}_k(\ell)$, we calculate the mean $\mu_k(\ell)$ and the deviation from the mean:
    \begin{equation}
        \vb{D}_k(\ell) = \vb{P}_k(\ell) - \mu_k(\ell).
    \end{equation}
    Next, we calculate the cumulative sum of the deviations as
    \begin{equation}
        Z_{i, k}(\ell) = \sum_{j = 1}^i D_{i, k}(\ell),
    \end{equation}
    for $i = 1, 2, \ldots, \ell$, and the range of each cumulative sum subseries as
    \begin{equation}
        R_k(\ell) = \max\vb{Z}_k(\ell) - \min\vb{Z}_k(\ell).
    \end{equation}
    Then, the mean of the rescaled ranges is calculated as
    \begin{equation}
        (R/S)_\ell = \expval{\frac{R_k(\ell)}{S_k(\ell)}} = \frac{1}{\kappa}\sum_{k = 1}^{\kappa}\frac{R_k(\ell)}{S_k(\ell)},
    \end{equation}
    where $S_k(\ell)$ is the standard deviation of the subseries $\vb{P}_k(\ell)$. We then repeat the process considering a different value of $\ell$ and the Hurst exponent is estimated assuming a power-law relation between the rescaled ranges and the length of the subseries $\ell$:
    \begin{equation}
        (R/S)_\ell \sim \ell^H,
    \end{equation}
    where $H$ is the Hurst exponent. The exponent $H$ is then estimated as the slope of the linear fit in the log-log plot of $(R/S)_\ell$ versus $\ell$, using the least squares method. Thus, $\ell \in [2, N / 2]$.

    \begin{figure}[t]
        \centering
        \includegraphics[width=\linewidth]{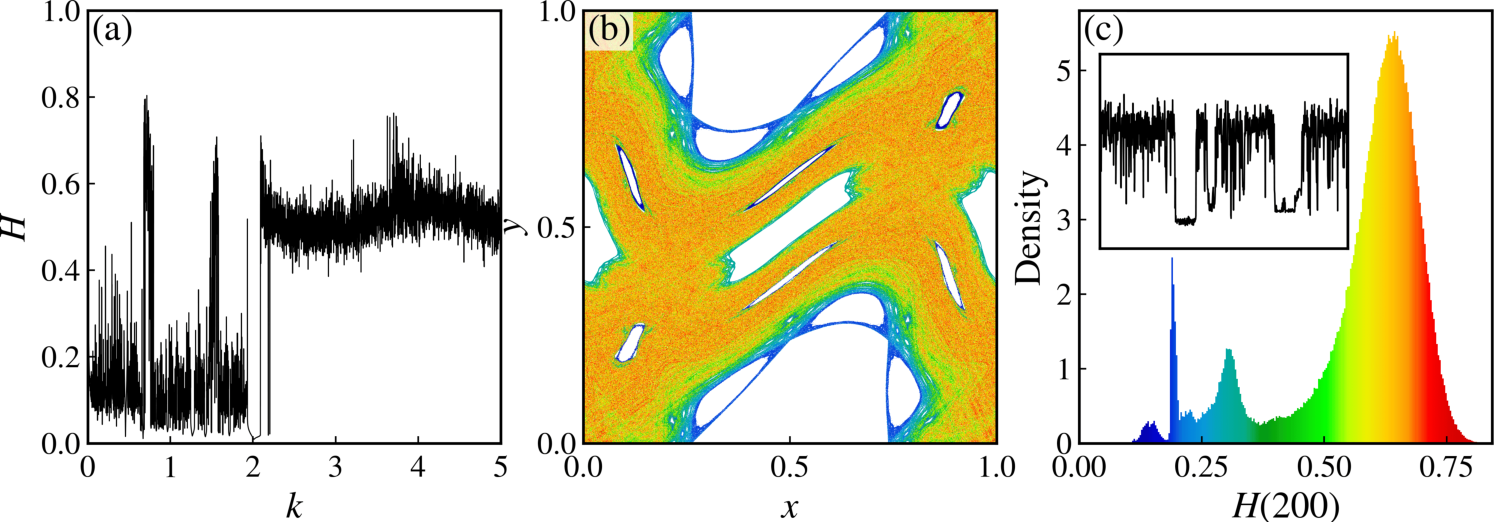}
        \caption{(a) Demonstration of the use of the \pyinline{hurst_exponent} method of the \pyinline{DiscreteDynamicalSystem} class of \pyinline{pynamicalsys} to calculate the Hurst exponent as a function of the parameter $k$ of the standard map [Eq.~\eqref{eq:stdmap}] and (b) and (c) demonstration of the use the \pyinline{finite_time_hurst_exponent} method of the \pyinline{DiscreteDynamicalSystem} class of \pyinline{pynamicalsys} to calculate the distribution of finite-time $H$ for the standard map with $k = 1.5$.}
        \label{fig:fig10}
    \end{figure}   

    This procedure considers only a $1$-dimensional time series. For a $d$-dimensional data, we simply apply it to each component of the time series, yielding a Hurst exponent vector $\vb{H} \in \mathbb{R}^d$. Depending on the problem we can analyze only one component of the Hurst exponent vector or consider its mean, for instance. The Hurst exponent can be calculated for a given discrete dynamical system using the \pyinline{hurst_exponent} method of the \pyinline{DiscreteDynamicalSystem} class from \pyinline{pynamicalsys}:
    \begin{lstlisting}[style=python]
obj.hurst_exponent(u, total_time, parameters=None, wmin=2, transient_time=None)        
    \end{lstlisting}
    Here, \pyinline{wmin} is the minimum length of each subseries and it is by default set to \pyinline{wmin=2}. This method returns the Hurst exponent vector $\vb{H}$ for higher-dimensional systems. To illustrate the calculation of the Hurst exponent, we consider the standard map [Eq.~\eqref{eq:stdmap}] and calculate the Hurst exponent as a function of the parameter $k$ with initial condition $(x, y) = 0.5, 0.25)$ [Fig.~\ref{fig:fig10}(a)]:
    \begin{lstlisting}[style=pycon]
>>> from pynamicalsys import DiscreteDynamicalSystem as dds
>>> ds = dds(model="standard map")
>>> u = [0.5, 0.25]  # Initial condition
>>> k_range = (0, 5, 5000)  # Interval in k
>>> k = np.linspace(*k_range)  # Create the k values
>>> total_time = 5000  # Total iteration time
>>> H = [ds.hurst_exponent(u, total_time, parameters=k[i]) for i in range(k_range[2])]
    \end{lstlisting}
    The Hurst exponent also distinguishes regularity from chaos accurately [cf.~Figs.~\ref{fig:fig5}(a) and~\ref{fig:fig9}(a)]. The regular regions yield a low value of $H$, while the chaotic regions exhibit a high value of $H$, just below $0.5$, which makes sense as $H = 0.5$ indicates a random walk and a chaotic trajectory does not fall within this category.
    
    Recently, Borin~\cite{Borin2024HurstTraps} has shown that the Hurst exponent is also an excellent tool for identifying and quantifying sticky orbits. By computing the Hurst exponent for a long chaotic trajectory of length $N$ (\pyinline{total_time}) in windows of size $n \ll N$ (\pyinline{finite_time}), it is possible to detect the different trappings around the hierarchical levels of the islands-around-islands structure. We use the \pyinline{finite_time_hurst_exponent} method of the \pyinline{DiscreteDynamicalSystem} class from \pyinline{pynamicalsys}:
    \begin{lstlisting}[style=python]
obj.finite_time_hurst_exponent(u, total_time, finite_time, parameters=None, wmin=2, return_points=False)        
    \end{lstlisting}
    Similarly to the other finite-time methods (Lyapunov and RTE), this method returns an array of $M = N / n$ rows with $d$ columns, where $d$ is the dimension of the system. The optional argument \pyinline{wmin} is the minimum length of each subseries and \pyinline{return_points} when set to \pyinline{True} tells the method to also return the initial conditions that generated the corresponding finite-time Hurst exponent value. The following code snippet illustrates the use of the \pyinline{finite_time_hurst_exponent} method for the calculation of the finite-time Hurst exponent distribution:
    \begin{lstlisting}[style=pycon]
>>> from pynamicalsys import DiscreteDynamicalSystem as dds
>>> ds = dds(model="standard map")
>>> u = [0.05, 0.05]  # Initial condition
>>> parameter = 1.5  # Parameter of the map
>>> total_time = 100000000  # Total iteration time
>>> finite_time = 200  # Finite time
>>> ftHE = ds.finite_time_hurst_exponent(u, total_time, finite_time, parameters=parameter)
>>> ftHE_avg = (ftHE[:, 0] + ftHE[:, 1]) / 2  # We use the average to calculate the distribution
    \end{lstlisting}

    The distribution of the finite-time Hurst exponent [Fig.~\ref{fig:fig10}(c)] is similar to the distribution of the finite-time RTE [Fig.~\ref{fig:fig9}(c)]. Both measures detect more than two modes and the Hurst exponent also distinguishes different trapping regions, as can be seen in Fig~\ref{fig:fig10}(b). The inset shows the same sharp drops in the values of $H(200)$, which yields the multi-modal distribution.    

    \section{Manifolds: the skeleton of the dynamics}
    \label{sec:manifolds}

    The stable and unstable manifolds are invariant geometric structures associated with saddle points of periodic orbits in dynamical systems. Given a discrete dynamical system $\vb{x}_{n + 1} = \vb{f}(\vb{x}_n)$, where $\vb{f}:\mathbb{R}^d\rightarrow\mathbb{R}^d$ is a smooth map, let $\vb{H}\in\mathbb{R}^d$ denote a hyperbolic fixed point (or a point on a hyperbolic periodic orbit) of $\mathbf{f}$. The stable manifold $W^s(\vb{H})$ is defined as the set of points $\mathbf{x}$ such that forward iterations under the map $\mathbf{f}$ asymptotically approach $\vb{H}$, i.e., $W^s(\vb{H}) = \left\{ \mathbf{x} \in \mathbb{R}^d \;\middle|\; \lim_{n \to \infty} \mathbf{f}^n(\mathbf{x}) = \vb{H} \right\}$.  Analogously, the unstable manifold $W^u(\vb{H})$ consists of all points whose backward iterates converge to $\vb{H}$, that is, $W^u(\vb{H}) = \left\{ \mathbf{x} \in \mathbb{R}^d \;\middle|\; \lim_{n \to -\infty} \mathbf{f}^n(\mathbf{x}) = \vb{H} \right\}$~\cite{YorkeBook}.
    
    Both manifolds are invariant under the dynamics of $\mathbf{f}$. The manifolds cross each other transversely an infinite number of times, which generates an infinite but countable set of saddle points immersed in the chaotic region, this set is called chaotic saddle~\cite{Pentek1995}. There are two types of intersections: homoclinic and heteroclinic. In the first case, the crossing is between the manifolds of the same hyperbolic point, while in the heteroclinic intersection, the manifolds of two distinct points cross each other~\cite{lichtenberg2013regular}. Due to the invariance of these manifolds under the dynamics, all forward and backward iterates of a homoclinic point also belong to both manifolds. Consequently, the existence of a single homoclinic point implies the existence of an infinite number of such points. In Hamiltonian systems, where phase space volume is conserved (Liouville’s theorem), the stable and unstable manifolds cannot intersect transversely just once. Instead, their intersections typically form intricate structures known as homoclinic tangles. 

    The relationship between the homoclinic tangle and the chaotic motion was given by Smale~\cite{Smale1967}, using the nonattractive set call Smale horseshoe $\Lambda$, a complete description of this can be found in~\cite{Ott2002}. The set $\Lambda$, associated with a Smale horseshoe, has the following properties: (i) includes a countable set of periodic orbits with arbitrarily large periods; (ii) an uncountable set of bounded aperiodic orbits; and (iii) at least one dense orbit is present. Birkhof~\cite{BirkhoffBook} and later Smale~\cite{Smale1963} demonstrated that every homoclinic point is an accumulation point of a family of infinitely many periodic orbits. Since the number of homoclinic points is infinite, it follows that in the neighborhood of each homoclinic point, there exist infinitely many periodic points. This implies the existence of an integer $n$ such that the $n$th iterate of the map exhibits a horseshoe structure $\Lambda$. Consequently, any orbit with an initial condition sufficiently close to a homoclinic tangle will have chaotic behavior.

    To numerically compute the stable and unstable manifolds of a given hyperbolic fixed point or hyperbolic periodic orbit, we first need to compute the eigenvectors of the Jacobian matrix. For two-dimensional maps, the Jacobian matrix evaluated at the hyperbolic fixed point has two eigenvalues, one larger than one and one less than one: $\abs{\lambda_1} > 1 > \abs{\lambda_2}$. The eigenvector with the largest eigenvalue, denoted by $\vb{v}_u$, represents the unstable direction, while the eigenvector with the smallest eigenvalue, denoted by $\vb{v}_s$, represents the stable direction. We select a large number of initial conditions uniformly distributed along the unstable eigenvector $\vb{v}_u$ and its negative counterpart $-\vb{v}_u$, within a distance $\delta \ll 1$ from the hyperbolic point. We then iterate these points forward in time, resulting in the unstable manifold of the hyperbolic fixed point (or hyperbolic periodic orbit). The stable manifold is obtained similarly. We distribute the initial conditions along the stable eigenvector $\vb{v}_s$ and its negative counterpart $-\vb{v}_s$ and iterate the initial conditions backward in time. This results in the stable manifold.

    Assuming we know the coordinates of the hyperbolic fixed points or at least on point on a hyperbolic periodic orbit, we can compute the stable and unstable manifolds using the \pyinline{manifold} method of the \pyinline{DiscreteDynamicalSystem} class from \pyinline{pynamicalsys}:
    \begin{lstlisting}[style=python]
obj.manifold(u, period, parameters=None, delta=1e-4, n_points=100, iter_time=100, stability="unstable")
    \end{lstlisting}
    In this case, \pyinline{u} represents the coordinates of the fixed point or periodic orbit, while \pyinline{period} is the orbit’s period. The argument \pyinline{delta} defines the distance from the fixed point where the initial conditions will be distributed. The arguments \pyinline{n_points} and \pyinline{iter_time} specify the number of points along the eigenvector and the number of iterations for each point, respectively. Finally, \pyinline{stability} indicates whether to calculate the stable or unstable manifold.

    However, before calculating the manifolds, we need to find and classify the fixed points and periodic orbits. The standard map [Eq.~\eqref{eq:stdmap}] has two fixed points: $(0, 0)$ and $(0.5, 0.0)$. We can analyze their stability analytically. The Jacobian matrix for the standard map is
    \begin{equation}
        J(x, y) = \mqty(1 + k\cos(2\pi x) & 1\\k\cos(2\pi x) & 1).
    \end{equation}
    For area-preserving maps, such as the standard map, the stability of the fixed points and periodic orbits can be estimated using their residue~\cite{Greene1968, Greene1979}:
    \begin{equation}
        R(x, y) = \frac{1}{4}\qty[2 - \Tr(J^p(x, y))],
    \end{equation}
    where $J(x, y)$ is the Jacobian matrix, $\Tr(\cdot)$ is the trace, and $p$ is the period of the orbit. An elliptic orbit has $R \in (0, 1)$, and a parabolic orbit has $R = 0$ or $R = 1$. The orbit is hyperbolic otherwise. Therefore, for the fixed point $(0, 0)$, we have $R(0, 0) = -k/4$, which means the fixed point is hyperbolic regardless of the value of $k$. For the other fixed point, we have $R(0.5, 0) = k/4$. Therefore, for $k \in (0, 4)$, the fixed point is elliptic. We can verify that using the \pyinline{classify_stability} method of the \pyinline{DiscreteDynamicalSystem} class from \pyinline{pynamicalsys}:
    \begin{lstlisting}[style=python]
obj.classify_stability(u, period, parameters=None) 
    \end{lstlisting}
    Here, \pyinline{u} is a list with the coordinates of the periodic orbit, \pyinline{parameters} is a list with the parameters of the system, and \pyinline{period} is the period of the periodic orbit. The below code snippet demonstrates the use of the \pyinline{classify_stability} method for the two fixed points we have discussed:
    \begin{lstlisting}[style=pycon]
>>> from pynamicalsys import DiscreteDynamicalSystem as dds
>>> ds = dds(model="standard map")
>>> period = 1 # Period of the orbit
>>> u = [0, 0] # Fixed point
>>> stability = ds.classify_stability(u, period, parameters=1.5)
>>> stability["classification"], stability["eigenvalues"]
('saddle', array([3.18614066+0.j, 0.31385934+0.j]))
>>> stability = ds.classify_stability(u, period, parameters=5.0)
>>> stability["classification"], stability["eigenvalues"]
('saddle', array([6.85410197+0.j, 0.14589803+0.j]))
>>> u = [0.5, 0] # Fixed point
>>> stability = ds.classify_stability(u, period, parameters=1.5)
>>> stability["classification"], stability["eigenvalues"]
('elliptic (quasi-periodic)', array([0.25-0.96824584j, 0.25+0.96824584j]))
>>> u = [0.5, 0]
>>> stability = ds.classify_stability(u, period, parameters=5.0)
>>> stability["classification"], stability["eigenvalues"]
('saddle', array([-2.61803399+0.j, -0.38196601+0.j]))
    \end{lstlisting}

\begin{figure}[t!]
        \centering
        \includegraphics[width=\linewidth]{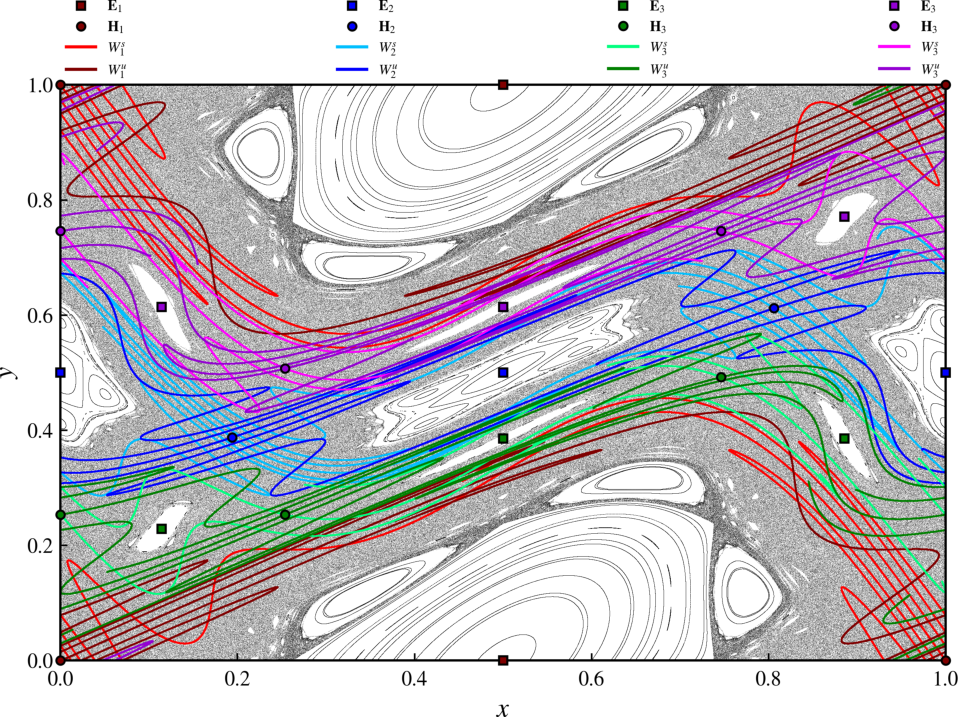}
        \caption{Demonstration of the use of the \pyinline{find_periodic_orbit} and \pyinline{manifold} methods of the \pyinline{DiscreteDynamicalSystem} class of \pyinline{pynamicalsys} using the standard map [Eq.~\eqref{eq:stdmap}] with $k = 1.5$ as an example.}
        \label{fig:manifolds}
    \end{figure}

    Once we have determined the stability of the fixed points, we can calculate the manifolds as follows:
    \begin{lstlisting}[style=pycon]
>>> saddle = [0, 0]
>>> period = 1
>>> k = 1.5
>>> n_points = 50000
>>> iter_time = 12
>>> wu = ds.manifold(saddle, period, parameters=k, n_points=n_points, iter_time=iter_time, stability="unstable")
>>> ws = ds.manifold(saddle, period, parameters=k, n_points=n_points, iter_time=iter_time, stability="stable")        
    \end{lstlisting}
    The manifolds of the hyperbolic fixed point are displayed in maroon (unstable) and red (stable) in Fig.~\ref{fig:manifolds}.

    Sometimes it is very difficult or even impossible to find the periodic orbits analytically, especially for higher periods. In this case, we can use the \pyinline{find_periodic_orbit} method of the \pyinline{DiscreteDynamicalSystem} class from \pyinline{pynamicalsys} to perform a two-dimensional scan of a region in phase space where a periodic orbit might be present:
    \begin{lstlisting}[style=python]
obj.find_periodic_orbit(grid_points, period, parameters=None, tolerance=1e-5, max_iter=1000, convergence_threshold=1e-15, tolerance_decay_factor=0.5, verbose=False, symmetry_line=None, axis=None)        
    \end{lstlisting}
    This method locates a periodic orbit within a specified region of phase space using an iterative grid refinement strategy. The input \pyinline{grid_points} is a three-dimensional array of shape \pyinline{(grid_size_x, grid_size_y, 2)} representing a mesh of initial conditions in phase space. The method identifies periodic points by evolving each initial condition for a fixed number of steps defined by the \pyinline{period} argument and checking whether the trajectory returns within a specified \pyinline{tolerance} of its starting point. This tolerance acts as a numerical threshold, effectively defining the neighborhood within which a return is considered periodic. If periodic points are detected, their minimum and maximum positions are used to define a new, smaller region of phase space. The grid is then refined within this updated region, and the process repeats. The refinement continues until either the position of the orbit converges to within a specified \pyinline{convergence_threshold} or the maximum number of iterations given by \pyinline{max_iter} is reached. At each iteration, the \pyinline{tolerance} is decreased according to the \pyinline{tolerance_decay_factor} argument and if convergence is detected earlier, the process halts before reaching the maximum iteration count. By default, the method does not print any information to the user. However, by setting \pyinline{verbose=True}, the method will print the convergence information at each refinement iteration.

    Even though this is an efficient method, the search can be improved by using the symmetries of the map to reduce the 2D search to finding the root of a function of one variable~\cite{Greene1979}. If a map, denoted by $M$, is reversible, then it can be written as a product of two involutions:
    \begin{equation}
        \label{eq:involutions}
        M = I_2 \circ I_1,
    \end{equation}
    where an involution is a map such that $I(I(\vb{x})) = \vb{x}$, i.e., $I^2 = I\circ I = 1$. The symmetry lines correspond to the fixed point sets of involutions, that is, the set of points where $I(\vb{x}) = \vb{x}$. For example, the standard can be written in the form of Eq.~\eqref{eq:involutions} using
    \begin{equation}
        \begin{aligned}
            I_1 &= \qty(-x, y + \frac{k}{2\pi}\sin\qty(2\pi x)),\\
            I_2 &= \qty(x - p, -p).
        \end{aligned}
    \end{equation}
    A point $(x^*, y^*)$ is on the symmetry line of $I_1$ if it satisfies $I_1(x^*, y^*) = (x^*, y^*)$. We get
    \begin{equation}
        \begin{aligned}
            &-x^* = x^*\bmod 1,\\
            &\sin\qty(2 \pi x^*) = 0.
        \end{aligned}
    \end{equation}
    Thus, the fixed set of $I_1$ is $\{(x, y)\,\mid\, x = 0\text{ or } 1/2\,\forall\,y\}$. These are the vertical lines at $x = 0$ and $x = 0.5$. Using $I_2$ we find the horizontal line at $y = 0$. Note, however, that these are not the only symmetry lines. By using different involutions that satisfy Eq.~\eqref{eq:involutions}, it is possible to find more symmetry lines~\cite{Pina1987}.
    
    Let us assume, for instance, that a symmetry line has the form $y(x) = f(x, \vb{p})$, where $\vb{p} = (p_1, p_2, \ldots)$ denotes the parameters of the system. Then, it is possible to define a function that takes on the coordinate $x$ and the system's parameters: \pyinline{y = symm_line(x, parameters)} and pass it to the \pyinline{find_periodic_orbit} method via \pyinline{symmetry_line=symm_line}. It is also necessary to define the axis of the symmetry line. In this case, \pyinline{axis=0}. If, however, the symmetry line has the form $x(y) = g(y, \vb{p})$, the procedure is analogous: the symmetry line function should be defined as \pyinline{x = symm_line(y, parameters)} and \pyinline{symmetry_line=symm_line} and \pyinline{axis=1}. 

    For instance, to find the elliptic periodic orbit of period 2 of the standard map for $k = 1.5$ (see Fig.~\ref{fig:fig1} for reference), we can perform the search along the vertical line $x = 0.0$:
    \begin{lstlisting}[style=pycon]
>>> symmetry_line = lambda y, parameters: 0.0 * np.ones_like(y)
>>> k = 1.5
>>> period = 2
>>> y_range = (0.4, 0.6, 1000)
>>> y = np.linspace(*y_range)
>>> tolerance = 2 / len(y) # Initial tolerance for period detection
>>> periodic_orbit = ds.find_periodic_orbit(y, period, parameters=k, tolerance=tolerance, symmetry_line=symmetry_line, axis=1, verbose=False)
>>> periodic_orbit
array([0. , 0.5])
>>> stability = ds.classify_stability(periodic_orbit, period, parameters=k)
>>> stability["classification"], stability["eigenvalues"]
('elliptic (quasi-periodic)', array([-0.125-0.99215674j, -0.125+0.99215674j]))
    \end{lstlisting}

    Now, for the hyperbolic periodic orbit, we know that it is somewhere in between the two period 2 islands (Poincaré-Birkhoff theorem). So instead of trying to find another symmetry line to find these points, we perform a two-dimensional search within the region $(x, y) \in [0.1, 0.3] \times [0.3, 0.55]$:
    \begin{lstlisting}[style=pycon]
>>> k = 1.5    
>>> period = 2
>>> grid_size = 1000
>>> tolerance = 2 / grid_size # Initial tolerance for period detection
>>> x_range = (0.1, 0.3, grid_size) # Limits of the rectangular region
>>> y_range = (0.3, 0.5, grid_size)
>>> x = np.linspace(*xrange) # Generate a grid of points in the rectangular region
>>> y = np.linspace(*yrange)
>>> X, Y = np.meshgrid(x, y) # Create a meshgrid of points in the rectangular region
>>> grid_points = np.empty((grid_size, grid_size, 2)) # 3D array of points in the rectangular region
>>> grid_points[:, :, 0] = X
>>> grid_points[:, :, 1] = Y
>>> periodic_orbit = ds.find_periodic_orbit(grid_points, period, parameters=k, tolerance=tolerance)
>>> periodic_orbit
array([0.19397649, 0.38795298])
>>> stability = ds.classify_stability(periodic_orbit, period, parameters=k)
>>> stability["classification"], stability["eigenvalues"]
('saddle', array([4.09176343+0.j, 0.24439341+0.j]))
    \end{lstlisting}

    The calculation of the manifolds is similar to the period 1 case and Figure~\ref{fig:manifolds} shows the fixed points and periodic orbits up to period three and also the stable and unstable manifolds of the saddles. The data of the figure has been generated using the methods we have discussed in this section. We have chosen, however, not to display all the code in this paper due to its size, and we refer the reader to the Supplementary Material for further details.

    \section{Escape analysis}
    \label{sec:escape}

    In this section, we discuss the escape dynamics in discrete dynamical systems using \pyinline{pynamicalsys}. In general, escape dynamics describes the statistical behavior of a collection of trajectories that leave a bounded region in phase space or escape through exits or holes present in the system. The analysis of the escape dynamics is essential to understanding the dynamical behavior of the system as it exhibits different statistical properties depending on the underlying dynamics. When two or more exits are present in the system, one can construct escape basins to analyze the uncertainty associated with the dynamical behavior. This approach is analogous to the basins of attraction in dissipative systems that have more than one attractor. Therefore, in general, we talk about escape basins in Hamiltonian systems, which preserve volume in phase space and thus cannot have attractors and basins of attraction in dissipative systems. This, however, does not prevent us from studying escape in dissipative systems.
    
    For open systems, such as the Hénon map~\cite{henon1976} or the Hénon-Heiles system~\cite{Henon1964}, the definition of escape is natural: the trajectory has escaped when it leaves toward infinity~\cite{Aguirre2001, Aguirre2003WadaSystems, Custdio2011, Dettmann2012, Vallejo2025}. In closed systems, however, it is necessary to introduce exits in the system. This is a classical approach, especially for Hamiltonian systems~\cite{Mugnaine2020, Nieto2024SystematicValues, Souza2024, RolimSales2024b, RolimSales2024c, SimileBaroni2025TransportMap}. When the dissipative system has more than one attractor or the open system has more than one exit, the corresponding basins are divided by a basin boundary that can be either a smooth curve or a fractal curve. Smooth boundaries are related to regular dynamics, whereas fractal boundaries are a classical characteristic of chaotic dynamics. Additionally, fractal boundaries decrease the predictability of the final state~\cite{Grebogi1983, McDonald1985, Grebogi1987}, i.e., to which basin the initial condition belongs. For a complete review of fractal boundaries in dynamical systems, we refer the reader to Ref.~\cite{Aguirre2009} and references therein.

    \subsection{Survival probability}

    The escape times, i.e., the time it takes for a trajectory to escape through one of the exits also tell us important information regarding the underlying dynamics. Given the escape time, we compute the survival probability, $P(n)$, that corresponds to the fraction of initial conditions that have not escaped until the $n$th iteration. It is defined as,
    \begin{equation}
        \label{eq:Survivel_prob}
        P(n) = \frac{N_{\text{surv}}(n)}{M},
    \end{equation}
    where $M$ is the total number of initial conditions and $N_{\text{surv}}$ is the number of initial conditions that have not escaped until the $n$th iteration. In a hyperbolic and strongly chaotic system, the survival probability decays exponentially as $P(n)\sim\exp{(-\kappa n)}$, where $\kappa$ is known as the escape rate. However, in Hamiltonian systems, the existence of trapping regions leads to a slower escape rate: instead of an exponential decay, then the decay can be a power-law~\cite{Altmann2009} or a stretched exponential~\cite{Dettmann2012}.
   
\begin{figure}[t]
        \centering
        \includegraphics[width=\linewidth]{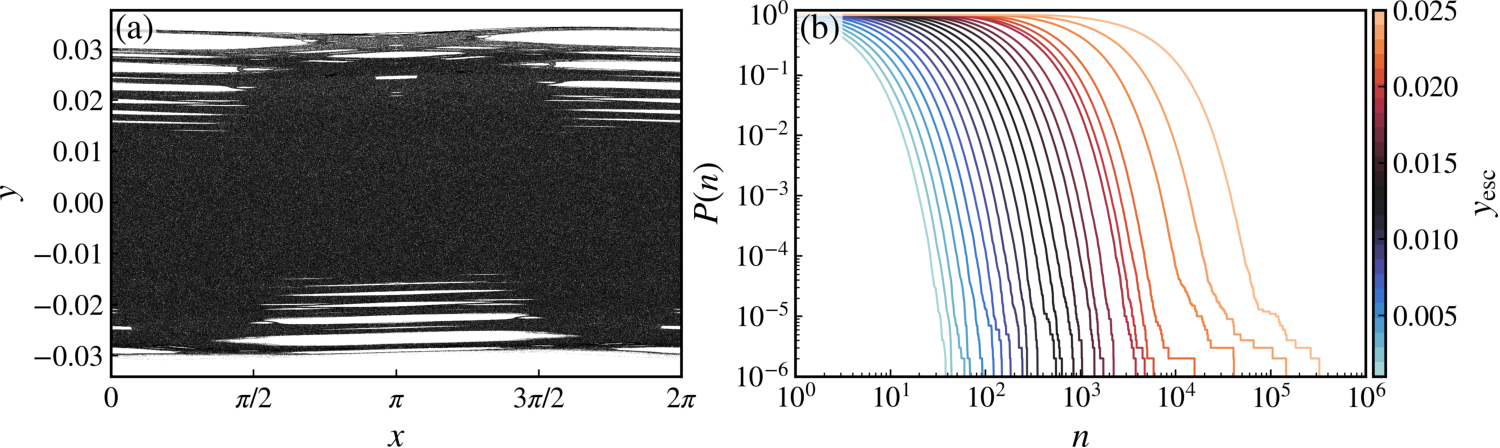}
        \caption{Demonstration of the use of the \pyinline{escape_analysis} with \pyinline{escape="exiting"} and \pyinline{survival_probability} methods of the \pyinline{DiscreteDynamicalSystem} class of \pyinline{pynamicalsys} for the Leonel map [Eq.~\eqref{eq:leonelmap}] for different escape regions defined by $y_{\mathrm{esc}}$.}
        \label{fig:survprob}
    \end{figure}

    To illustrate this feature, we consider the Leonel map~\cite{deOliveira2010, Leonel2011}, defined as
    \begin{equation}
        \label{eq:leonelmap}
        \begin{aligned}
            y_{n + 1} &= y_n + \epsilon\sin\qty(x_n),\\
            x_{n + 1} &= x_n + \frac{1}{\abs{y_{n + 1}}^\gamma} \bmod{2\pi},
        \end{aligned}
    \end{equation}
    where $\epsilon > 0$ is the nonlinearity parameter and $\gamma > 0$ controls the speed of the divergence of the $x$ coordinate in the limit $y \rightarrow 0$. This mapping has the interesting feature of exhibiting chaotic regions for small, but nonzero, perturbation values ($\epsilon \ll 1$) due to the divergent behavior of the second term in the $y$ equation. For small values of $y$, $1/\abs{y}^\gamma\rightarrow\infty$ and $x_{n + 1}$ and $x_n$ becomes uncorrelated, thus generating chaotic behavior. As $y$ increases, becomes slower and slower, and regular regions can be found in phase space [Fig.~\ref{fig:survprob}(a)]. The transition from integrability ($\epsilon = 0$) to non-integrability ($\epsilon \neq 0$) has been investigated for this system and researchers have shown that the transition is characterized by a second-order phase transition. Additionally, the diffusion of chaotic orbits in the system is scaling invariant, i.e., the diffusion can be characterized by a homogeneous function of the parameters $\epsilon$ and $\gamma$. By introducing symmetric exits, located at the horizontal lines $y = \pm y_{\text{esc}}$, one can show that the survival probability is also scaling invariant as long as the phase space region $(x, y) \in [0, 2\pi] \times [-y_{\mathrm{esc}}, y_{\mathrm{esc}}]$ does not contain any stability islands~\cite{Borin2023}.

    The escape analysis can be done using the \pyinline{escape_analysis} method from the \pyinline{DiscreteDynamicalSystem} class of \pyinline{pynamicalsys}:
    \begin{lstlisting}[style=python]
obj.escape_analysis(u, max_time, exits, parameters=None, escape="entering", hole_size=None)
    \end{lstlisting}
    Here, the argument \pyinline{max_time} determines the maximum iteration time to check for escape. If this time is reached and the trajectory has not escaped, the simulation is stopped. The argument \pyinline{exits} defines the exits of the system. The shape of this argument depends on the \pyinline{escape} argument. If \pyinline{escape="entering"}, it means that the trajectories escape upon entering a predefined region, i.e., by reaching a hole in the system. However, if the escape happens when trajectories leave a predefined region, such as in our current example, then this argument should be set to \pyinline{escape="exiting"}. The argument \pyinline{hole_size} defines the size of the hole when \pyinline{escape="entering"}.

    In this section, we are going to focus on \pyinline{escape="exiting"} and we discuss the other case in Sec.~\ref{sec:escbasins}. The argument \pyinline{exits} defines the boundaries of the $d$-dimensional phase-space box and has the format \pyinline{exits = [[x_ini, x_end], [y_ini, y_end], [z_ini, z_end]]}, where each sublist specifies the lower and upper bounds along one coordinate axis. The \pyinline{escape_analysis} method returns two values: \pyinline{escape_side}, which indicates the side through which the trajectory escaped (with \pyinline{-1} meaning no escape was detected), and \pyinline{escape_time}, which is the time it took for the trajectory to escape (equal to \pyinline{max_time} if no escape occurred). In general, escape can happen through any side of the box: 0 (1) corresponds to the left (right) side, 2 (3) to the bottom (top), and the pattern continues in higher dimensions. In the specific example considered here, the $x$ direction is periodic, so if we define \pyinline{exits = [[0, 2 * np.pi], [-y_esc, y_esc]]}, escape can only occur in the $y$ direction. Escapes through the $x$ boundaries (i.e., \pyinline{escape_side=0} for $x < 0$ or \pyinline{escape_side=1} for $x > 2\pi$) are excluded due to the periodicity in $x$.

    The following code snippet demonstrates the use of the \pyinline{escape_analysis} method. It calculates the escape and survival probability for different $y_{\mathrm{esc}}$ for random initial conditions defined in the interval $(x, y) \in [0, 2\pi]\times[-1\times10^{-14}, 1\times10^{-14}]$ [Fig.~\ref{fig:survprob}(b)]:
    \clearpage
    \begin{lstlisting}[style=pycon]
>>> from pynamicalsys import DiscreteDynamicalSystem as dds
>>> ds = dds(model="leonel map")
>>> ds.info["parameters"]
["eps", "gamma"]
>>> eps, gamma = 1e-3, 1.0  # Define the parameters
>>> parameters = [eps, gamma]
>>> max_time = 1000000  # Maximum time
>>> num_ic = 1000000  # Number of initial conditions
>>> np.random.seed(13)  # Seed for reproducibility
>>> x_range = (0, 2 * np.pi, num_ic)  # Limits in x for the initial conditions
>>> y_range = (-1e-14, 1e-14, num_ic)  # Limits in y for the initial conditions
>>> x = np.random.uniform(*x_range)  #  Create the random initial initial conditions
>>> y = np.random.uniform(*y_range)
>>> y_esc = np.logspace(np.log10(1e-3), np.log10(0.025), 25)  # Define the escape region
>>> x_esc = (0, 2 * np.pi)
>>> sp, times = [], []  # Empty list to store the survival probability and times
>>> for i in range(len(y_esc)):
...     exit = np.array([[x_esc[0], x_esc[1]], [-y_esc[i], y_esc[i]]])
...     escape = [ds.escape_analysis([x[j], y[j]], max_time, exit, parameters=parameters, escape="exiting") for j in range(num_ic)]
...     escape = np.array(escape, dtype=np.int32)
...     time, survival_probability = ds.survival_probability(escape[:, 1], escapes[i, :, 1].max())
...     sp.append(survival_probability)
...     times.append(time)
    \end{lstlisting}

    The exponential decay is evident for almost all survival probability curves. The last three curves, i.e., the largest $y_{\text{esc}}$ values, exhibit a small power-law tail for large values of $n$. This is characteristic of systems that exhibit the stickiness effect.

    \subsection{Escape basins}
    \label{sec:escbasins}

    Due to the strong sensitivity to initial conditions, chaotic systems often exhibit escape basins with an intertwined pattern and fractal basin boundaries. We exemplify this with a two-dimensional, area-preserving nontwist map, which describes the advection of passive scalars~\cite{Weiss1991,Pierrehumbert1991}. The map is defined by the following equations:
    \begin{equation}
        \label{eq:weissmap}
        \begin{aligned}
            y_{n+1} & = y_n - k \sin(x_n), \\
            x_{n+1} & = x_n + k \left( y_{n+1}^2 - 1 \right)\bmod 2\pi, 
        \end{aligned}
    \end{equation}
    where, $k > 0$ is the nonlinearity parameter. This map is called nontwist due to the violation of the twist condition $\pdv*{x_{n + 1}}{y_n} \neq 0$. Indeed, by calculating this derivative, we obtain
    \begin{equation}
        \pdv{x_{n + 1}}{y_n} = 2ky_{n + 1}.
    \end{equation}
    The twist condition is violated for $y_{n + 1} = 0$. Nontwist systems exhibit nonmonotonic rotation number profiles, which leads to the phenomenon of degeneracy, i.e., two or more distinct stability islands with the same rotation number. For the Weiss map, due to the quadratic dependence of $x_{n + 1}$ on $y_n$, there are two sets of islands with the same rotation number~\cite{Souza2023b}.

    Additionally, nontwist systems exhibit a robust transport barrier, called the shearless curve. The shearless curve corresponds to a local extremum of the rotation number profile and it prevents global transport: it divides the phase space into two distinct and unconnected domains. Numerous studies on the breakup of the shearless curve have been done and we refer the reader to Refs.~\cite{del-Castillo-Negrete1996, delcastillonegrete1997, Mathias2019, Szezech2009, Mugnaine2018, Szezech2012, Mugnaine2024, SimileBaroni2025TransportMap, Grime2025} and references therein for more details on two-dimensional, area-preserving nontwist systems.

\begin{figure}[t]
        \centering
        \includegraphics[width=\linewidth]{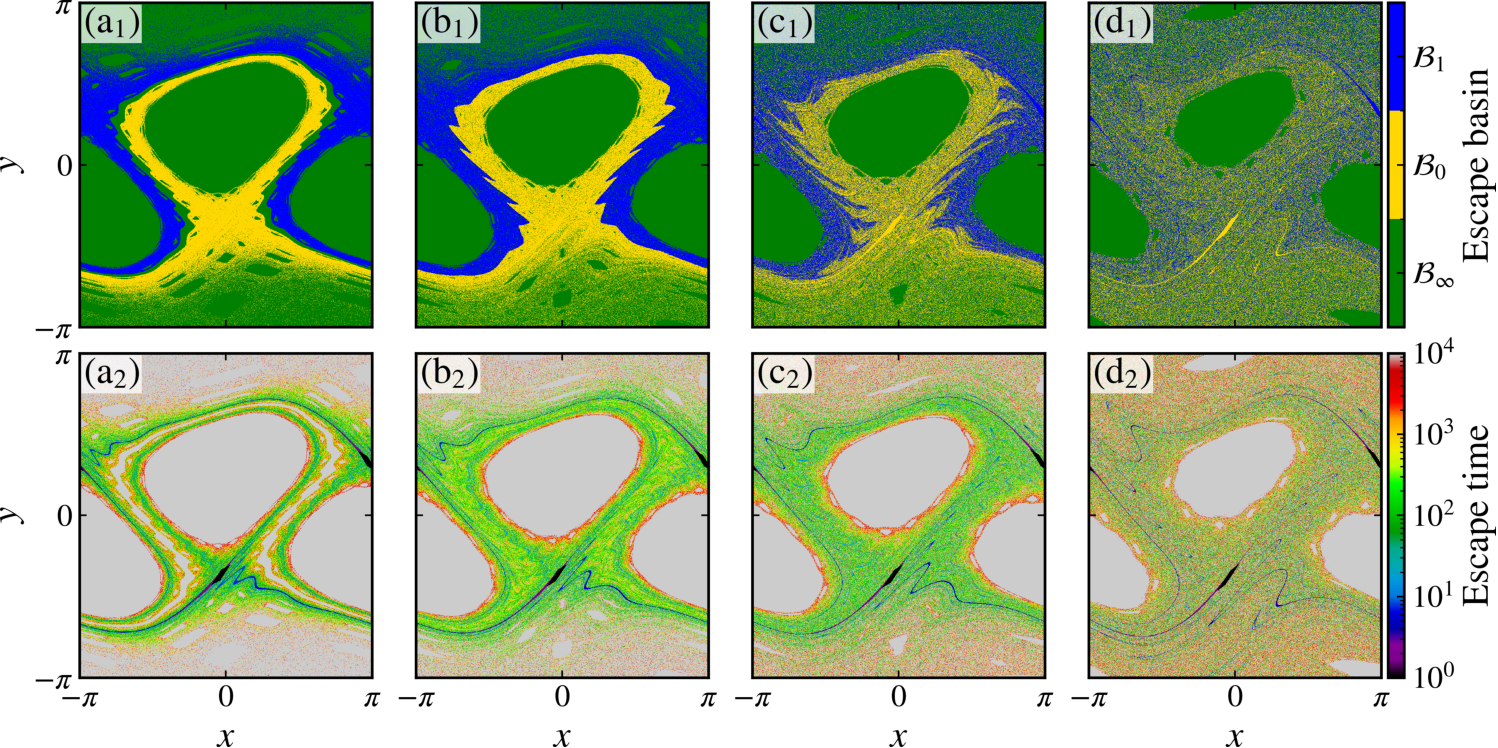}
        \caption{Demonstration of the use of the \pyinline{escape_analysis} method with \pyinline{escaping="entering"} for the Weiss map [Eq.~\eqref{eq:weissmap}] with different values of $k$, namely, (a) $k = 0.50$, (b) $k = 0.55$, (c) $k = 0.60$, and (d) $k = 0.70$.}
        \label{fig:escape_basins}
    \end{figure}

    Contrary to the approach we took on the previous section, now we introduce two holes of width $0.2$ in the phase space of the system, centered at the points $(x, y) = (0.0, -1.1)$ and $(x, y) = (\pi - 0.1, 1.0)$~\cite{Souza2023b}. The \pyinline{exits} argument now corresponds to the centers of the holes and the \pyinline{hole_size} theirs the widths. Also, we must modify the \pyinline{escape} argument: \pyinline{escape="entering"}. But first, we need to define the mapping function as this system is not built-in within the \pyinline{DiscreteDynamicalSystem} class:
    \begin{lstlisting}[style=pycon]
>>> from pynamicalsys import DiscreteDynamicalSystem
>>> from numba import njit
>>> @njit
>>> def weiss_map(u, parameters):
...     k = parameters[0]
...     x, y = u
...     y_new = y - k * np.sin(x)
...     x_new = (x + k * (y_new ** 2 - 1) + np.pi) % (2 * np.pi) - np.pi
...     return np.array([x_new, y_new])
>>> ds = dds(mapping=weiss_map, system_dimension=2, number_of_parameters=1)
    \end{lstlisting}
    To generate the escape basins in this case, we consider a $1000\times1000$ grid of initial conditions uniformly distributed on $(x, y) \in [-\pi, \pi]^2$ and iterate each one of them up to $10^4$ times. The following code snippet illustrates the calculation of the escape basins for four different values of $k$, namely, $k = 0.5$, $k = 0.55$, $k = 0.60$, and $k = 0.70$ (Fig.~\ref{fig:escape_basins}):
    \begin{lstlisting}[style=pycon]
>>> import itertools
>>> centers = [[0.0, -1.1], [np.pi - 0.1, 1.0]]
>>> hole_size = 0.2
>>> grid_size = 1000
>>> ks = [0.5, 0.55, 0.60, 0.70]
>>> total_time = 10000
>>> x_range = (-np.pi, np.pi, grid_size)
>>> y_range = (-np.pi, np.pi, grid_size)
>>> X = np.linspace(*x_range)
>>> Y = np.linspace(*y_range)
>>> escapes = np.zeros((len(ks), grid_size, grid_size, 2))
>>> for i, k in enumerate(ks):
...     escape = []
...     for x, y in itertools.product(X, Y):
...         escape.append(ds.escape_analysis([x, y], total_time, centers, parameters=k, hole_size=size_exit))
...     escape = np.array(escape).reshape(grid_size, grid_size, 2)
...     escapes[i, :, :, :] = escape
    \end{lstlisting}

\begin{figure}[t]
        \centering
        \includegraphics[width=\linewidth]{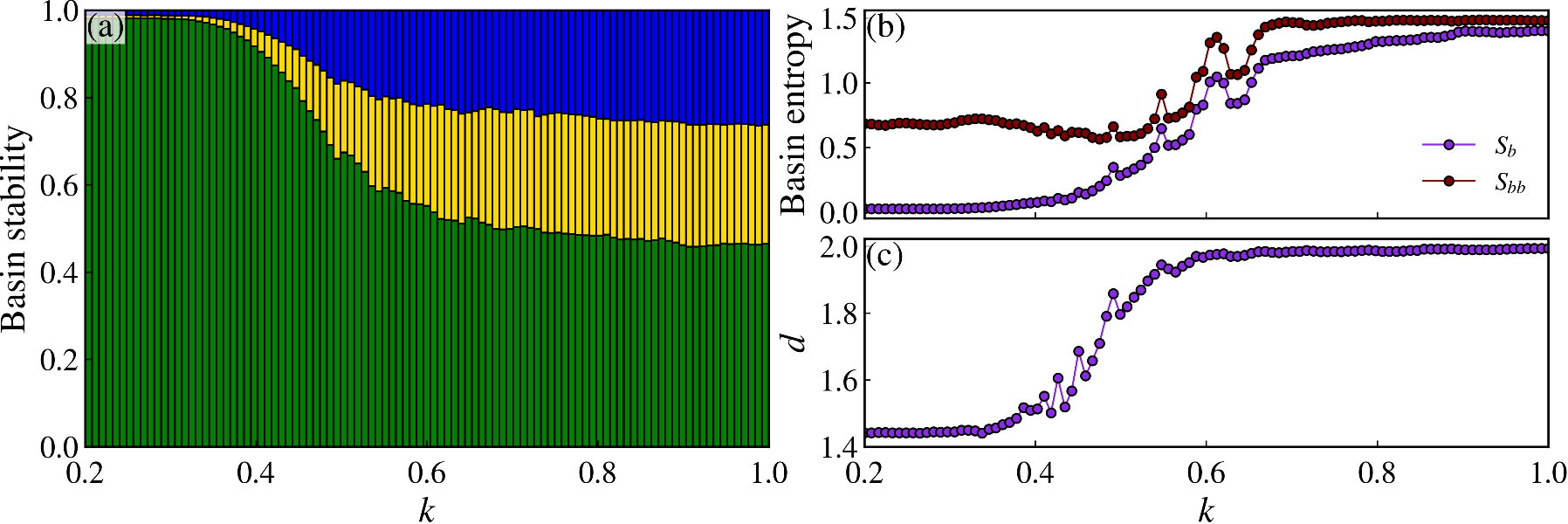}
        \caption{(a) The basin stability and (b) and (c) the demonstration of the use of the \pyinline{basin_entropy} and \pyinline{uncertainty_fraction} methods of the \pyinline{BasinMetrics} class of \pyinline{pynamicalsys}.}
        \label{fig:escape_quant}
    \end{figure}

    In this case, the first output of the \pyinline{escape_analysis} method can only be -1, 0, or 1. If the initial condition has not reached one of the exits until $10^4$ iterations, it returns -1 and we color the point green. If, however, the initial condition has reached the first (second) exit, it returns 0 (1) and we color the point yellow (blue). The first row of Fig.~\ref{fig:escape_basins} corresponds to the escape basins while the second row corresponds to the escape times. For $k = 0.5$, the shearless curve is still present in the phase space of the system, i.e., the blue and yellow basins do not mix. As we increase the perturbation, the shearless curve is broken and we observe an intermixing of the basins. To understand the effect of the parameter $k$ on the size of the basins, we can compute the basin stability~\cite{Menck2013}, which is simply the relative proportion of each basin, Fig~\ref{fig:escape_quant}(a). The stability of the green basin, which corresponds to the points that do not escape, for being in trapping regions or inside KAM islands, with the increase of the nonlinearity parameter these islands are broken, having the areas reduced leading to a decrease of the stability of this basin. The stability of the yellow and blue basins increases with the decrease of the regular regions, both basins present similar values for the basin stability, which indicate a strong mixing and intertwined basins. 
    
    To quantify the uncertainty of the final state caused by the fractal structure of the basin boundaries, we compute the basin entropy and boundary entropy as well as the dimension of the boundary using the uncertainty fraction method~\cite{Grebogi1983, McDonald1985, Grebogi1987}. Regarding the basin entropy, the method was introduced by Daza et al.~\cite{Daza2016} and it consists of dividing the escape basin, characterized by the presence of $N_e$ distinguishable asymptotic states, into a fine mesh of $N\times N$ boxes. Each box contains a set of initial conditions that leads to a certain asymptotic state, which in our case can be escaping through one of the exits or never escaping. We label these asymptotic states from 1 to $N_e$. For each box $i$, we associate a probability $p_{ij}$ of the asymptotic state $j$ to be present in the box and define the Shannon entropy of the $i$th box as
    \begin{equation}
        S_i = - \sum_{j=1}^{n_i}p_{ij}\log{p_{ij}},
    \end{equation}
    where $n_i \in [1, N_e]$ is the number of asymptotic states present in the box. The total basin entropy is obtained by averaging over all boxes:
    \begin{equation}
        \label{eq:S_b}
        S_b = \frac{1}{N^2}\sum_{i=1}^{N^2}S_i.
    \end{equation}

    The $S_b$ quantity is a measure of the complexity of the escape basin as a whole, with higher values indicating more complex basins. This methodology also allows us to compute the uncertainty of the final state associated with the basin boundary. To do this, we follow the same procedure to obtain Eq.~\eqref{eq:S_b}, but considering only the $N_b$ boxes that contain more than one asymptotic state, i.e., those that intersect multiple basins. Then the basin boundary entropy is
    \begin{equation}
        S_{bb} = \frac{1}{N_b}\sum_{i = 1}^{N^2}S_i = \frac{NS_b}{N_b}.
    \end{equation}
    A sufficient but not necessary condition for the boundary to be fractal is $S_{bb} > \log 2$. If we consider the logarithm in the base $2$, this condition becomes $S_{bb} > 1$.
    
    Fractal boundaries are those with a non-integer value for their dimension. In our case, we have a two-dimensional basin and a smooth boundary is characterized by $d = 1$, while a fractal boundary exhibits $d > 1$. We compute the dimension of the boundary using the uncertainty fraction method~\cite{Grebogi1983, Grebogi1987, McDonald1985}. Similarly to the basin entropy method, given an escape basin and an uncertainty $\epsilon$, for each point, we test whether small perturbations along each coordinate axis by $\epsilon$ remain in the same basin. Specifically, given a point $(x_i, y_i)$, we evaluate whether the perturbed points $(x_i \pm \epsilon, y_i)$ and $(x_i, y_i \pm \epsilon)$, one at a time, converge to the same asymptotic state as the reference point. If at least one of these four perturbed points belongs to a different basin, the reference point is classified as $\epsilon$-uncertain. The uncertainty fraction $f(\epsilon)$ is defined as the ratio of $\epsilon$-uncertain points to the total number of points in the basin. By varying $\epsilon$, we obtain the dependence of $f(\epsilon)$ on $\epsilon$.

    For a smooth boundary, $f(\epsilon)\sim\epsilon$, whereas for fractal boundaries the uncertainty fraction scales with $\epsilon$ as a power law: $f(\epsilon) \sim \epsilon^\alpha$, where $\alpha$ is the uncertainty exponent. The uncertainty exponent, $\alpha$, and the dimension of the boundary, $d$, are related through the following equation~\cite{McDonald1985}:
    \begin{equation}
        d = D - \alpha,
    \end{equation}
    where $D$ is the dimension of the basin. In our case, $D = 2$ and a fractal boundary is characterized by $\alpha\in(0, 1)$.

    Both entropies and the uncertainty fraction can be calculated using the \pyinline{basin_entropy} and \pyinline{uncertainty_fraction} methods, respectively, of the \pyinline{BasinMetrics} class of \pyinline{pynamicalsys}. To instantiate this class, you simply pass as an argument the two-dimensional array representing the basin you wish to quantify:
    \begin{lstlisting}[style=pycon]
>>> basin = np.random.randint(1, 4, size=(1000, 1000))  # Example basin
>>> from pynamicalsys import BasinMetrics
>>> bm = BasinMetrics(basin)
    \end{lstlisting}
    The signatures of the methods are as follows:
    \begin{lstlisting}[style=python]
    obj.basin_entropy(n, log_base=np.e)
    obj.uncertainty_fraction(x, y, epsilon_max=0.1, epsilon_min=None, n_eps=100)
    \end{lstlisting}
    In the \pyinline{basin_entropy} method, the argument \pyinline{n} specifies the size of each box in the $N \times N$ grid that covers the basin, and \pyinline{log_base} sets the base of the logarithm used in the entropy calculation. This method returns a list containing the values of $S_b$ and $S_{bb}$.
    
    In the \pyinline{uncertainty_fraction} method, \pyinline{x} and \pyinline{y} are two-dimensional arrays that define the coordinates of the basin and must have the same shape as the basin. The arguments \pyinline{epsilon_max} and \pyinline{epsilon_min} specify the maximum and minimum values of the uncertainty $\epsilon$, in the same units as \pyinline{x} and \pyinline{y}. If \pyinline{epsilon_min} is not provided, the method automatically determines an appropriate value based on the resolution of the basin. For instance, if the basin has a resolution of $0.001$, setting \pyinline{epsilon_min} below this value is not meaningful, and in general, it is recommended to omit this argument. The final argument, \pyinline{n_eps}, determines the number of uncertainty values sampled between \pyinline{epsilon_min} and \pyinline{epsilon_max}. This method returns two arrays: the sampled values of $\epsilon$ and the corresponding values of the uncertainty fraction $f(\epsilon)$. The uncertainty exponent, and consequently the dimension, can be obtained by performing a least-squares fit of $\log \epsilon$ versus $\log f(\epsilon)$.
    
    In Figs.~\ref{fig:escape_quant}(b) and~\ref{fig:escape_quant}(c), we show the dependence of the basin entropy, $S_b$, and basin boundary entropy $S_{bb}$, and the dimension $d$ on the parameter $k$. Immediately we notice that the boundary is fractal, i.e., $d > 1$, for all values of $k$ while the fractality condition of $S_{bb} > \log_22$ is only satisfied for $k > 0.6$. This is consistent with our expectations, as the latter condition is sufficient but not necessary for fractality, as previously discussed. We also note that, with a few exceptions where $S_b$, $S_{bb}$, and $d$ oscillate, all quantities increase with $k$. This is also consistent with our expectations. As the nonlinearity parameter increases, the size of the stability islands diminishes and the basins become more and more mixed (Fig.~\ref{fig:escape_basins}). Moreover, for $k \rightarrow 1$, the dimension $d$ tends to $2$ and $S_b$ and $S_{bb}$ become closer and closer. This is an indication of riddled basins~\cite{Daza2022}, i.e., basins in which every neighborhood of a point in one basin contains points belonging to other basins, making the system extremely sensitive to initial conditions. In terms of the uncertainty fraction, a value of $d \approx 2$ implies $\alpha \approx 0$. In other words, no matter how much we reduce the uncertainty $\epsilon$, the number of uncertain points does not decrease.
    


    \section{Conclusions}
    \label{sec:concl}

    In this paper, we have introduced \pyinline{pynamicalsys}, an open-source Python module for the analysis of discrete dynamical systems. The module implements a variety of methods to analyze and quantify the dynamical behavior of discrete dynamical systems. These include trajectory and bifurcation diagram computation, Lyapunov exponent estimation, the smaller alignment index (SALI) and the linear dependence index (LDI), and other indicators of chaotic behavior. Additionally, it provides tools for periodic orbit detection and the computation of their invariant manifolds, as well as escape analysis and basin quantification. All methods are built on top of Numpy and Numba, ensuring high performance and efficiency. The \pyinline{DiscreteDynamicalSystem} class comes with several built-in models ready to use, however, it is not limited to the built-in ones. The definition of a custom mapping function is extremely easy and straightforward.

    We have provided a description, literature review, and mathematical description of the principal methods and classes of \pyinline{pynamicalsys}'s module. Additionally, the complete documentation is available in Ref.~\cite{documentation}, with a more in-depth discussion and beginner-friendly language along with the API (Application Programming Interface) reference. The Jupyter notebook used to generate and plot all the data used in this paper is available in the Supplementary Material as well as in Ref.~\cite{githubsup}. Even tough \pyinline{pynamicalsys} currently only supports discrete dynamical systems, we are already working on a new version that includes continuous-time systems as well, and we are committed to keep implementing new features and including them in new versions of \pyinline{pynamicalsys}. 

    Even though there has been a paradigm shift in the scientific community regarding the public sharing of code and data~\cite{Besan2021}, many researchers remain reluctant~\cite{Gomes2022}, and some do not comply with their own published data sharing statements~\cite{Gabelica2022}. This has led to reproducibility issues in many scientific publications~\cite{Baker2016, Miyakawa2020}. Therefore, we hope that \pyinline{pynamicalsys} contributes to making research in nonlinear dynamics more accessible and reproducible, and helps shift the prevailing culture of withholding code and data.

    A similar package to the \pyinline{pynamicalsys} module can be found in the Julia programming language, the \pyinline{DynamicalSystems.jl} package~\cite{Datseris2018, DatserisParlitz2022}, and there are some other open-source projects related to numerical methods in nonlinear dynamics research, such as the \pyinline{pyunicorn}~\cite{pyunicorn}, \pyinline{ordpy}~\cite{Pessa2021}, \pyinline{tisean}~\cite{Hegger1999}, and \pyinline{powerlaw}~\cite{Alstott2014}.

    \section*{Supplementary Material}

    See the Supplementary Material for the Jupyter notebook (or its respective PDF file) for the codes to generate and plot all the data used in this article. The notebooks also include measures of CPU times of the methods described in this paper.

    \section*{Code availability}

    The source code to reproduce the results presented in this paper is freely available in the GitHub repository at Ref.~\cite{github}.

    \section*{Acknowledgments}

    This work was supported by the São Paulo Research Foundation (FAPESP, Brazil), under Grant Nos.~2019/14038-6, 2021/09519-5, 2023/08698-9, 2024/09208-8, 2024/03570-7, 2024/14825-6, and 2024/05700-5, and by the National Council for Scientific and Technological Development (CNPq, Brazil), under Grant Nos.~301318/2019-0, 304616/2021-4, 309670/2023-3, and 304398/2023-3.


%

  \end{document}